\documentclass[preprint,showpacs,preprintnumbers,groupedaddress,%
amsmath,amssymb,floatfix,pra]{revtex4}

\usepackage{epsfig}
\usepackage{epsf}
\usepackage{amsmath}
\usepackage{amsfonts}
\usepackage{amssymb}
\usepackage{bm}
\usepackage{bbm}
\usepackage{nicefrac}
\usepackage{color}
\usepackage{pifont}
\usepackage{graphicx}
\usepackage{epsfig}
\usepackage{latexsym}

\def\e{   {\rm e}   }
\def\arctantwo#1#2{\text{arctan}_2\;(#1,#2)}

\newcommand{\vect}[1]{\bm{#1}}
\newcommand{\ud}{d}
\newcommand{\Tr}{\textrm{Tr}}
\newcommand{\kb}{k_b}
\newcommand{\kc}{k_c}
\newcommand{\kbtilde}{\tilde{k}_b}
\newcommand{\kctilde}{\tilde{k}_c}
\newcommand{\hatkb}{\hat{k}_b}
\newcommand{\hatkc}{\hat{k}_c}
\newcommand{\boldkb}{\bm{k}_b}
\newcommand{\boldkc}{\bm{k}_c}
\begin{document}

\title{Correlated two--photon emission by transitions of Dirac--Volkov states
in intense laser fields: QED predictions}

\author{Erik L\"otstedt}
\email{Erik.Loetstedt@mpi-hd.mpg.de}
\affiliation{Max-Planck-Institut f\"{u}r Kernphysik,
Postfach 103980, 69029 Heidelberg, Germany}
\altaffiliation[Present address: ]{Department of Chemistry, 
School of Science, The University of Tokyo, 7-3-1 Hongo, 
Bunkyo-ku, Tokyo 113-0033, Japan}

\author{Ulrich D. Jentschura}
\affiliation{Department of Physics,
Missouri University of Science and Technology,
Rolla, Missouri 65409-0640, USA}
\affiliation{Institut f\"ur Theoretische Physik, Universit\"at Heidelberg,
Philosophenweg 16, 69120 Heidelberg, Germany}

\begin{abstract}
In an intense laser field, 
an electron may decay by emitting a pair of photons.
The two photons emitted during the process,
which can be interpreted as a laser-dressed 
double Compton scattering, remain entangled in a 
quantifiable way: namely, the so-called concurrence
of the photon polarizations gives a gauge-invariant 
measure of the correlation of the hard gamma rays.
We calculate the differential rate and
concurrence for a backscattering setup of
the electron and photon beam, employing
Volkov states and propagators for the
electron lines, thus accounting nonperturbatively 
for the electron-laser interaction. The nonperturbative
results are shown to differ significantly compared
to those obtained from the usual double Compton scattering.
\end{abstract}


\maketitle
%

%
%
\section{Introduction}
\label{intro}

\begin{figure}[thb]
\begin{center}
\includegraphics[width=1\columnwidth]{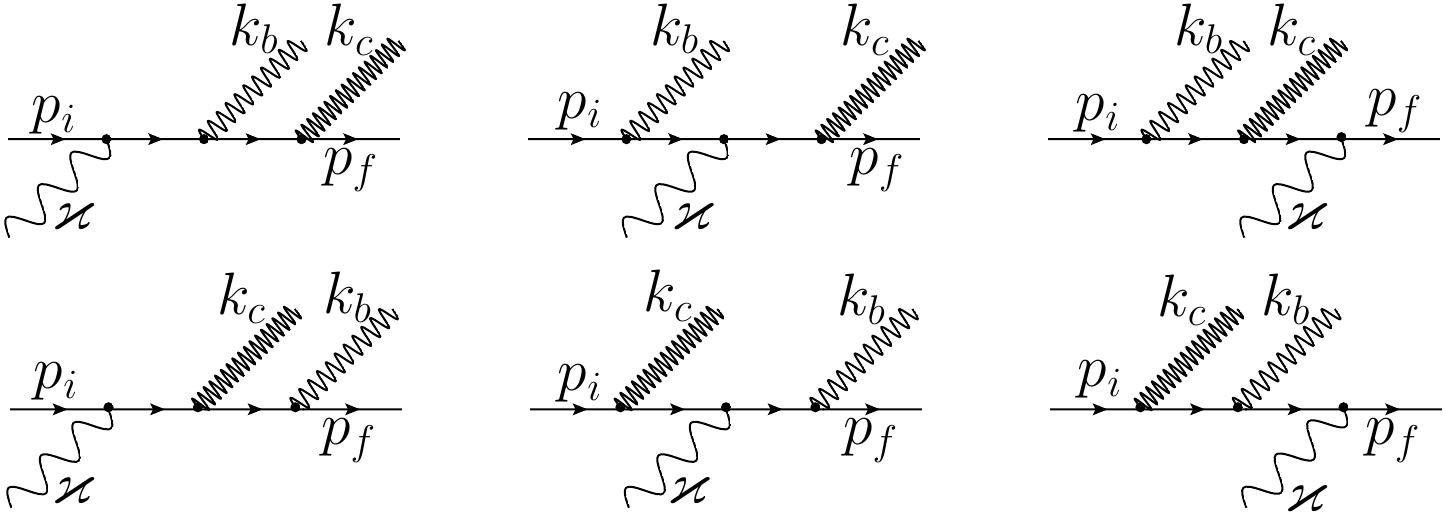}
\caption{The 6 Feynman diagrams contributing to (ordinary) double Compton
scattering (initial electron four momentum $p_i$, final four momentum 
$p_f$). The electron line is denoted by a single customary 
fermion line. The frequencies of the two emitted photons with wave four vectors
$\kb$ and $\kc$ may be different and thus may be their wavelengths;
this is explicitly indicated in the panel. \label{FG}}
\end{center}
\end{figure}

\begin{figure}[t!]
\begin{center}
\includegraphics[width=1\columnwidth]{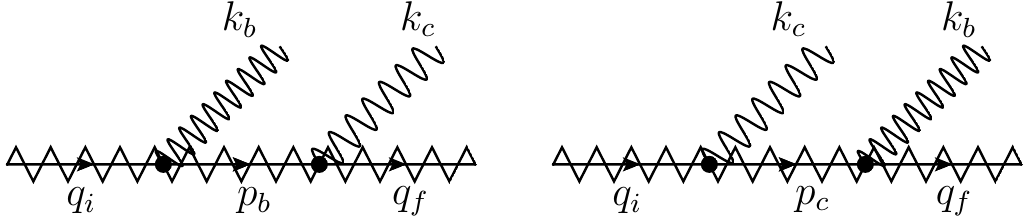}
\caption{Feynman diagram for the two-photon decay of a Dirac--Volkov 
state. The electron line is dressed by the laser field and denoted 
by a zigzag laser photon superimposed on the fermion line.
\label{fig_Feynman_diag}}
\end{center}
\end{figure}

In perturbative double Compton scattering 
\cite{Eliezer1946,MandlSkyrme1952,Bell2008}, an incoming 
photon interacts with an electron, and two photons are
emitted. This process, which is represented by the 
Feynman diagrams in Fig.~\ref{FG}, can be described 
by perturbative quantum electrodynamics (QED) and 
requires no other special theoretical input. Experimental 
evidence ranges from the first measurements more than 
50 years ago~\cite{Cavanagh1952,TheusBeach1957,McGieBradyKnox1966} 
to the more recent~\cite{SeSaGhu1988,SaDeSiGhu1999,SaSaSi2006,SaSiSa2008}. 
However,
if the emission process takes place inside an intense laser field, 
then the physics changes, and the electron line is 
dressed by multiple interactions with the laser field
(see Fig.~\ref{fig_Feynman_diag}).
The emission of two photons is a purely quantum 
process which cannot be described by classical radiation
theory~\cite{Mel1972}. An exception is encountered only 
for the case of the sequential emission of 
two quanta which occurs when the intermediate propagator 
hits a resonance pole, given by a resonant 
Dirac--Volkov state. In that case, to which we will return to later
in the paper, the diagrams in Fig.~\ref{fig_Feynman_diag}
 break 
apart into two distinctive blocks for the emission of the 
two photons. 

The most interesting geometry for the process is 
the backscattering case, where a relativistic electron 
counterpropagates against an intense laser beam 
of comparatively low frequency 
(on the order of a few eV). 
In ordinary Compton scattering, the electron 
is usually assumed to be at rest, and the 
scattering of a highly energetic photon is considered.
Because the kinematics is inverted in the 
backscattering case, one sometimes refers to this 
scenario as ``inverse Compton scattering''.
During the emission,
the electron interacts with the laser field via an 
arbitrary number of interactions 
(see Fig.~\ref{fig_Feynman_diag}); the process can be described by 
fully laser-dressed Dirac--Volkov propagators
\cite{ReEb1966,LoJeKe2007}.
So, we may refer to the process depicted in 
Fig.~\ref{fig_Feynman_diag} as ``inverse laser-dressed double Compton
backscattering.'' 

Note that for a single-photon Compton backscattering, 
the highest photon energy attainable is $4 \gamma^2 \omega$,
where $\omega$ is the laser photon energy and
$\gamma$ is the Lorentz factor of the incoming electron. 
For a defined scattering geometry, the energy 
of the emitted photon thus is uniquely defined, and it 
coincides with the energy of the emitted classical (Larmor)
radiation in the specified direction provided 
the laser photon energy is much smaller than 
the electron mass and 
Lorentz boost factors are taken into account.
If the electron absorbs $n$ laser photons during 
laser-dressed single Compton scattering, the 
energy maximum changes to $4 n \gamma^2 \omega/(1+\xi^2)$, 
where the laser intensity parameter $\xi$ is defined in 
Eq.~\eqref{xiEbar} below ($\xi^2$ is proportional to the 
laser intensity). When two photons are emitted in laser-dressed
double Compton scattering, their 
maximum energy sum is limited by 
$\omega_1 + \omega_2 \le 4 n \gamma^2 \omega/(1+\xi^2)$.
As we will show, it is possible to designate energy and 
angular regions in which the 
double scattering process dominates over single scattering, 
which is crucial for an experimental 
verification~\cite{BroMarBinColEva2008,Thirolfetal2009}.

Interestingly, as noted in \cite{ScScHa2006,ScScHa2008,ScMai2009},
the two photons emitted during the process are entangled
because of the quantum nature of the process.
In order to quantify the entanglement, the emission directions 
of the two quanta cannot be used with good effect,
because they represent continuous variables in three dimensions.
However, the polarization components of the two photons along the emission 
lines can be uniquely decomposed in a two-dimensional space 
composed of unit vectors (effectively a one-dimensional space),
and measured independently. Triggering on simultaneous 
two-photon events, one can then measure the entanglement
quantitatively: an appropriate measure is the 
so-called concurrence \cite{Wootters1998,Cirone2005269}
which measures the polarization entanglement of the two quanta. 

The usual double Compton scattering, which involves 
the absorption of only one laser photon, has a rate which 
is proportional to the square of the laser four-vector 
amplitude, i.e., proportional to its intensity.
Therefore, we may refer to the single scattering process
as the ``linear'' process. 
With rising laser intensity, the rate
deviates from the simple linear intensity 
dependence, it becomes more and more indispensable to 
include higher-order effects, and the process becomes nonlinear.

In order to bring the current investigation into perspective,
we would like to mention other work 
performed in connection with two-photon emission from free electrons:
indeed, a pair of photons may be produced by electrons accelerated 
by any kind of external field. Probably the most well known
process of this kind is  double bremsstrahlung 
\cite{Baier1981293,AltQuar1985,VenGavMaq1987,Hippler1991,%
KahLiuQuar1992,%
Korol1997,DonFlo1998,KryMaMaSta2002,Korol2006}, but also 
photon pair creation in a magnetic field 
\cite{ZhuNi1973,SoVoletal1976,FoKho2003},
and in a crossed field \cite{MoRi1974} has been considered.
The process under investigation in this paper is complementary to those 
 mentioned above, and may provide for better control
of the properties of the produced photons by adjusting the 
laser parameters.

This paper is organized as follows.
In Sec.~\ref{formulation}, we discuss the formulation in terms
of a laser-dressed (``nonperturbative'') QED formalism.
We then continue, in Sec.~\ref{perturbative},
with a comparison of the predictions of the fully relativistic,
nonperturbative theory to the relativistic, but perturbative
(in the laser field) theory of double Compton scattering. In 
particular, we extend the discussion given 
in Ref.~\cite{LoJe2009_2} to also include 
circularly polarized laser fields.
In Sec.~\ref{angular}, we study the angular correlation and the 
entanglement of the emitted photons, in the nonperturbative 
formalism. Finally, conclusions are drawn in 
Sec.~\ref{conclu}. Throughout the paper, we use relativistic natural
units such that $\hbar=c=1$, and a space-time
metric $g^{\mu\nu} = {\rm diag}(1, -1, -1, -1)$. Scalar products
of four-vectors are written as $p_\mu q^\mu=p\cdot q$ for 
two four-vectors $p$ and $q$. The gamma matrices are written as $\gamma^\mu$, 
and their contraction with a four-vector $p$ as $\hat{p}=\gamma\cdot p$.

%
%
\section{Formulation of the QED theory}
\label{formulation}

\subsection{Notation}

The electron mass is denoted by $m$, and the electron charge by $e=-|e|$.
The laser wave vector points in the negative $x^3$ direction 
[with the space-time coordinate $x^\mu=(x^0,\bm{x})=(x^0,x^1,x^2,x^3)$],
\begin{equation}
\varkappa^\mu=(\omega,\bm{\varkappa})=\omega \, (1,0,0,-1),
\end{equation}
and the laser four-vector potential, modeled as 
a monochromatic plane wave, for linear polarization
is
\begin{equation}\label{vector_pot_lin}
A_{\text{lin}}^\mu(\phi)=a^\mu \cos\phi,\qquad \phi=\varkappa\cdot x,
\end{equation}
with $a\cdot \varkappa =0$, $a^\mu=|a|(0,1,0,0)$.
For circular polarization we have instead,
\begin{equation}\label{vector_pot_circ}
A_{\text{circ}}^\mu(\phi)=a_1^\mu\cos\phi+
a_2^\mu\sin\phi,
\end{equation}
with $a_1\cdot a_2 = 0$, $a_2^2=a_1^2$, $a_1^\mu=|a_1|(0,1,0,0)$,
$a_2^\mu=|a_1|(0,0,1,0)$.
The laser intensity parameter $\xi$ is defined as 
\begin{equation}\label{defxilin}
\xi=\frac{-e}{m}\sqrt{\frac{-a^2}{2}},
\end{equation}
for linear, and
\begin{equation}\label{defxicirc}
\xi=\frac{-e\sqrt{-a_1^2}}{m} =
\frac{-e}{m}\sqrt{\frac{-a_1^2 - a_2^2}{2}}. 
\end{equation}
for circular polarization. For a consistent comparison of linear and circular
polarization, one should compare at the same value of $\xi$,
which corresponds to the same laser intensity.
The parameter $\xi$ relates to
the  root-mean-square
electric field amplitude $\bar{E}$ like
\begin{equation}\label{xiEbar}
\xi=\frac{-e\bar{E}}{m\omega},
\end{equation}
and can be said to be the relativistic (inverse of the) Keldysh parameter:
$\xi<1$ corresponds to the multiphoton regime of relativistic 
laser-matter interaction, where the coupling
to the laser field is perturbative, and $\xi>1$ is commonly 
referred to as the tunneling, or nonperturbative regime.
The quantum  parameter $\chi$ \cite{Ri1985}, which 
in general determines the magnitude of quantum effects such 
as $e^+e^-$ pair creation, spin effects etc, is defined as
\begin{equation}\label{chidef}
\chi=\xi \, \frac{p_i\cdot \varkappa}{m^2},
\end{equation}
where $p_i$ is the initial momentum of the electron 
[see Eq.~\eqref{defpi}].
If we compute $\chi$ in the rest frame of the electron, where
$p_i=(m,\bm{0})$, then 
\begin{equation}
\chi=\xi \frac{\omega}{m} =\frac{\bar{E}}{E_{\textrm{crit}}},
\end{equation}
where $E_{\textrm{crit}}=m^2/|e|$ is the critical (Schwinger) field. Thus,
$\chi$ is the amplitude of the electrical field of the laser compared to the 
critical field in the rest frame of the electron. 
The relation to laser intensities follows from the formula
\begin{equation}\label{intensity}
I=\xi^2\left(\frac{\omega}{m}\right)^2 I_{\textrm{crit}},
\end{equation}
where $I_{\textrm{crit}}=2.3\times 10^{29}$ W/cm$^2$ is the critical intensity, 
corresponding to $\chi=1$ in the lab frame.
For most of our examples, we use $\omega=2.5$ eV (optical laser), which
corresponds to $I=5.5\times 10^{18}$ W/cm$^2$ for $\xi=1$ and 
$I=2.2\times 10^{19}$ W/cm$^2$ for $\xi=2$; the laser field
here is strong but manifestly sub-critical. Note that even in the case of 
a relativistic (Lorentz factor $\gamma=10^3$) electron beam as considered later in the examples 
(see Secs. \ref{gauge} and \ref{perturbative}), the laser field remains
sub-critical in the rest frame of the electron, since 
$\chi=\xi p_i\cdot \varkappa/m^2\approx 10^{-2} 
\ll 1$ with the parameters chosen.

The initial electron four-momentum is (we assume the electron 
to be counterpropagating with respect to the laser field, i.e.,
moving in the positive $x^3$-direction):
\begin{align}\label{defpi}
p_i =& \; 
(E_i,\bm{p}_i)=(E_i,0,0,\sqrt{E_i^2-m^2}), 
\nonumber\\
q_i =& \; p_i+
\xi^2\frac{m^2}{2\varkappa\cdot p_i}\,\varkappa=(Q_i,\bm{q}_i) \,,
\end{align}
which is valid for both circular and linear polarization. 
The final electron four-momentum is
\begin{equation}\label{defpf}
p_f = \; (E_f,\bm{p}_f),\quad q_f=p_f+
\xi^2\frac{m^2}{2\varkappa\cdot p_f}\, \varkappa=(Q_f,\bm{q}_f).
\end{equation}
The 
four-vector $q_{i,f}$ introduced in Eqs.~\eqref{defpi}, \eqref{defpf} 
is the 
average momentum of a laser-dressed electron \cite{Ri1985},
with corresponding average mass $m_\ast$, 
\begin{equation}
q_f^2 = \; 
q_i^2=m_\ast^2=m^2(1+\xi^2).
\end{equation}
The electron spinors are used in the following form:
\begin{equation}
u_r(p)=\sqrt{\frac{E+m}{2m}}
\left(\begin{array}{c}
\delta_{r1}\\\delta_{r2}\\
\frac{1}{E+m}\bm{\sigma}\cdot \bm{p} \binom{\delta_{r1}}{\delta_{r2}}
\end{array}\right),
\end{equation}
with the standard vector $\bm{\sigma}$ being 
composed of the (Pauli) $2 \times 2$ spin matrices. With this 
convention, the spinors are normalized according 
to $u_r^\dagger(p)\gamma^0 u_r(p)=\bar{u}_r u_r=1$.
For an electron moving in the $x^3$-direction, $r=1$ corresponds
to a right-handed electron, and $r=2$ to a left-handed electron.

The Volkov states \cite{Ri1985}, 
solutions of the Dirac equation with an external laser field
\begin{equation}
\left(i\hat{\partial}-m-e\hat{A}\right)\Psi=0,
\end{equation}
read for linear polarization [see Eq.~\eqref{vector_pot_lin}]
\begin{align}
\label{Volkov_lin}
\Psi_{p,r}(x) =& \; \sqrt{\frac{m}{QV}} \, \sum_{s=-\infty}^{\infty}
\left[A_0(s,\alpha,\beta)+\frac{e\hat{\varkappa}\hat{a}}
{2\varkappa\cdot p}A_1(s,\alpha,\beta)
\right] 
\nonumber\\[2ex]
& \times u_r(p) \e^{-i(q+s\varkappa)\cdot x},
\end{align}
where
\begin{equation}
\alpha = \frac{e \, a\cdot p}{\varkappa\cdot p}, 
\qquad 
\beta=\frac{e^2 \, a^2}{8\varkappa\cdot p}.
\end{equation}
Here, the generalized Bessel function \cite{Re1962,KoKlWi2006} 
is defined as
\begin{equation}
A_k(n,\alpha,\beta) = \;\frac{1}{2\pi}
\int_0^{2\pi} \cos^k\theta \, 
{\rm e}^{in\theta -i\alpha\sin\theta +i\beta\sin2\theta} \,
\ud\theta,
\end{equation}
with $k\ge 0$, from which follows
$A_{k>0}(n,\alpha,\beta)= \; \tfrac{1}{2}\left[
A_{k-1}(n+1,\alpha,\beta)+A_{k-1}(n-1,\alpha,\beta)\right]$.

For circular polarization [see Eq.~\eqref{vector_pot_circ}] we have,
\begin{align}\label{Volkov_cir}
\Psi_{p,r}(x) =& \;
\sqrt{\frac{m}{QV}}\sum_{s=-\infty}^{\infty}
\left[J_s(\bar{\alpha}) \, \e^{is\varphi } +
\frac{e \, \hat{\varkappa} \, \hat{a}_1}{2\varkappa\cdot p} 
J_s^+(\bar{\alpha},\varphi) \right.
\nonumber\\[2ex]
& \; \left. +\frac{e \, \hat{\varkappa} \, \hat{a}_2}{2\varkappa\cdot p} 
J_s^-(\bar{\alpha},\varphi)
\right] u_r(p) \e^{-i(q-s\varkappa)\cdot x}.
\end{align}
Here 
\begin{equation}
\bar{\alpha} =
\sqrt{\alpha_1^2+\alpha_2^2},\quad \alpha_1=\frac{ea_1\cdot p}{\varkappa\cdot p},
\quad \alpha_2=\frac{ea_2\cdot p}{\varkappa\cdot p},
\end{equation}
and
\begin{equation}
\varphi=\arctantwo{\alpha_2}{-\alpha_1}  \,.
\end{equation}
The $\arctantwo{\cdot}{\cdot}$ functions is defined as
\begin{align}\label{defarctan2}
\arctantwo{y}{x}&=\arctan\left(\frac{y}{x}\right)
\quad\text{if }x>0, \nonumber\\
\arctantwo{y}{x}&=\pi+\arctan\left(\frac{y}{x}\right)
\quad\text{if }x<0,
\end{align}
the usual Bessel functions are denoted by $J_n(\alpha)$, and
\begin{equation}\begin{split}
J_s^+(\alpha,\varphi)&=\frac{1}{2}\left[J_{s-1}(\alpha)\e^{i(s-1)\varphi}+
J_{s+1}(\alpha)\e^{i(s+1)\varphi}\right],\\
J_s^-(\alpha,\varphi)&=\frac{1}{2i}\left[J_{s-1}(\alpha)\e^{i(s-1)\varphi}-
J_{s+1}(\alpha)\e^{i(s+1)\varphi}\right] \,.
\end{split}
\end{equation}
Note the normalization factor in Eqs.~\eqref{Volkov_lin}
and~\eqref{Volkov_cir}:
the volume $V$ comes with the wave function, and not with the spinor $u(p)$.

\begin{figure}[thb]
\begin{center}
\includegraphics[width=0.8\columnwidth]{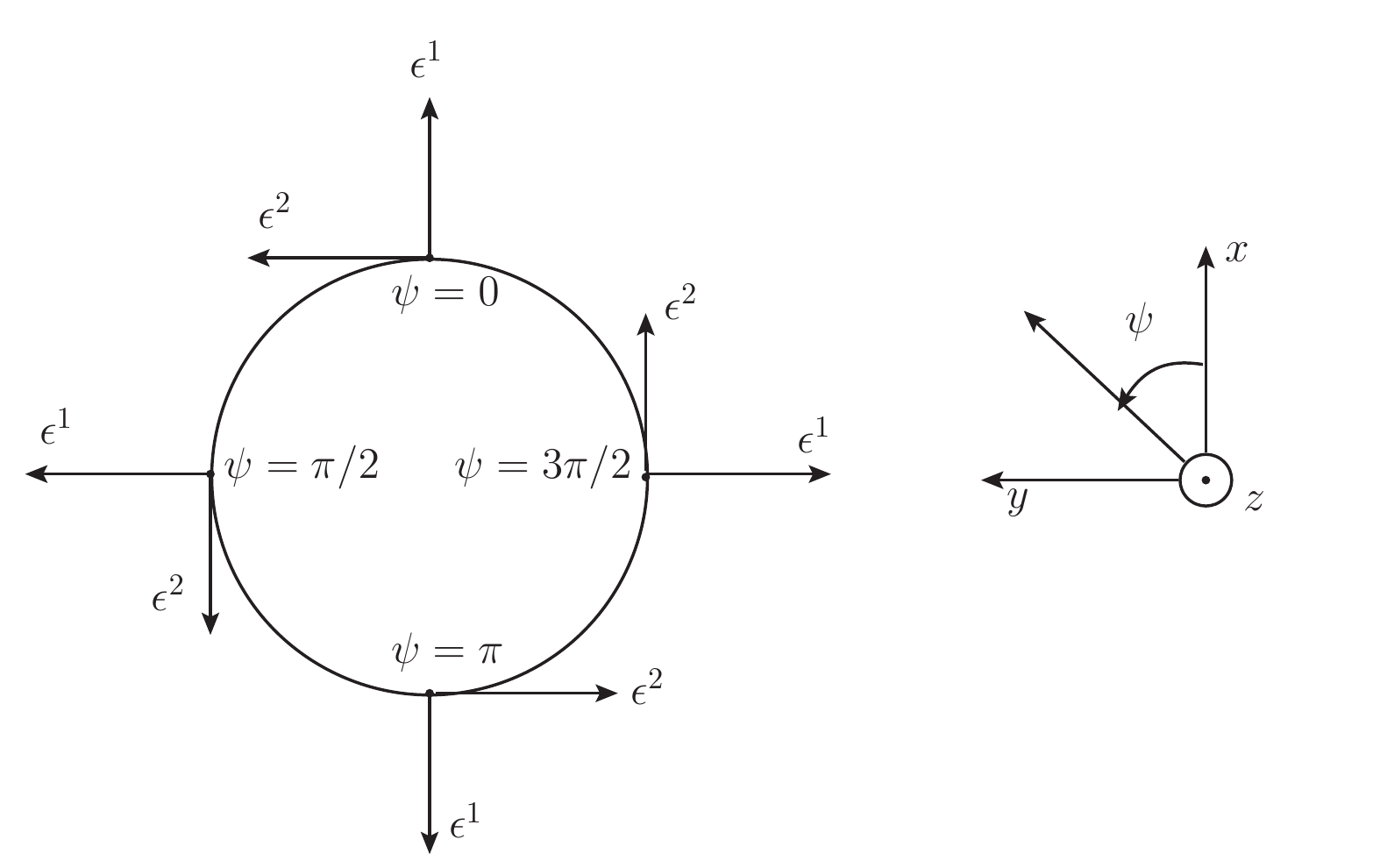}
\caption{Polarization vectors according to 
Eq.~\eqref{polvecs} for fixed, small $\theta \approx 0$, i.e.,~for 
a photon propagating in the positive $z$ direction.
For example, we have at $\psi=\pi/2$ the two vectors
$\epsilon^1=(0,0,\cos\theta, -\sin\theta)\approx(0,0,1,0)$ and 
$\epsilon^2=(0,-1,0,0)$.
\label{polv}}
\end{center}
\end{figure}

The propagation four-vectors of the two emitted photons are denoted by
\begin{subequations}
\begin{align}
\kb =& \; (\omega_b,\boldkb) = 
\omega_b \, \kbtilde 
\nonumber\\
=& \;
\omega_b \, (1, \, \sin\theta_b\cos\psi_b, \, \sin\theta_b\sin\psi_b,
\cos \theta_b),
\\
\kc =& \; (\omega_c,\boldkc) = 
\omega_c \, \kctilde
\nonumber\\
=& \;
\omega_c(1, \, \sin\theta_c\cos\psi_c, \, \sin\theta_c\sin\psi_c,
\cos \theta_c),
\end{align}
\end{subequations}
$\psi$ measuring the azimuth and $\theta$ measuring the 
polar angle. As a basis for the two  
polarization four-vectors $\epsilon_b$ and $\epsilon_c$ of the two 
emitted photons, we take
\begin{align}
\label{polvecs}
\epsilon_b^1=& \; 
\left(0, \, \cos\theta_b \, \cos\psi_b, \,
\cos\theta_b \, \sin\psi_b, \, -\sin\theta_b\right),
\nonumber\\
\epsilon_b^2=& \; 
\left(0, \, -\sin\psi_b, \, \cos\psi_b, \, 0 \right)
\nonumber\\
\epsilon_c^1 =& \; \left(0, \, \cos\theta_c \, \cos\psi_c, \,
\cos\theta_c \sin\psi_c, \, -\sin\theta_c \right)\,,
\nonumber\\
\epsilon_c^2=& \; \left(0, \, -\sin\psi_c, \, \cos\psi_c, \, 0\right) \,.
\end{align}
As an aid to the discussion, Fig.~\ref{polv} illustrates the 
direction of the polarization
vectors for small polar angle 
$\theta$ and different values of $\psi$.
Alternatively, the polarization can be expressed in a helicity basis 
according to
\begin{equation}
\label{polvecs_hel}
\begin{split}
&\epsilon^R_b=\frac{1}{\sqrt{2}}(\epsilon_b^1+i\,\epsilon_b^2),\qquad
\epsilon^L_b=\frac{1}{\sqrt{2}}(\epsilon_b^1-i\,\epsilon_b^2),\\
&
\epsilon^R_c=\frac{1}{\sqrt{2}}(\epsilon_c^1+i\,\epsilon_c^2),\qquad
\epsilon^L_c=\frac{1}{\sqrt{2}}(\epsilon_c^1-i\,\epsilon_c^2).
\end{split}\end{equation}

\begin{figure}[thb]
\begin{center}
\includegraphics[width=0.8\columnwidth]{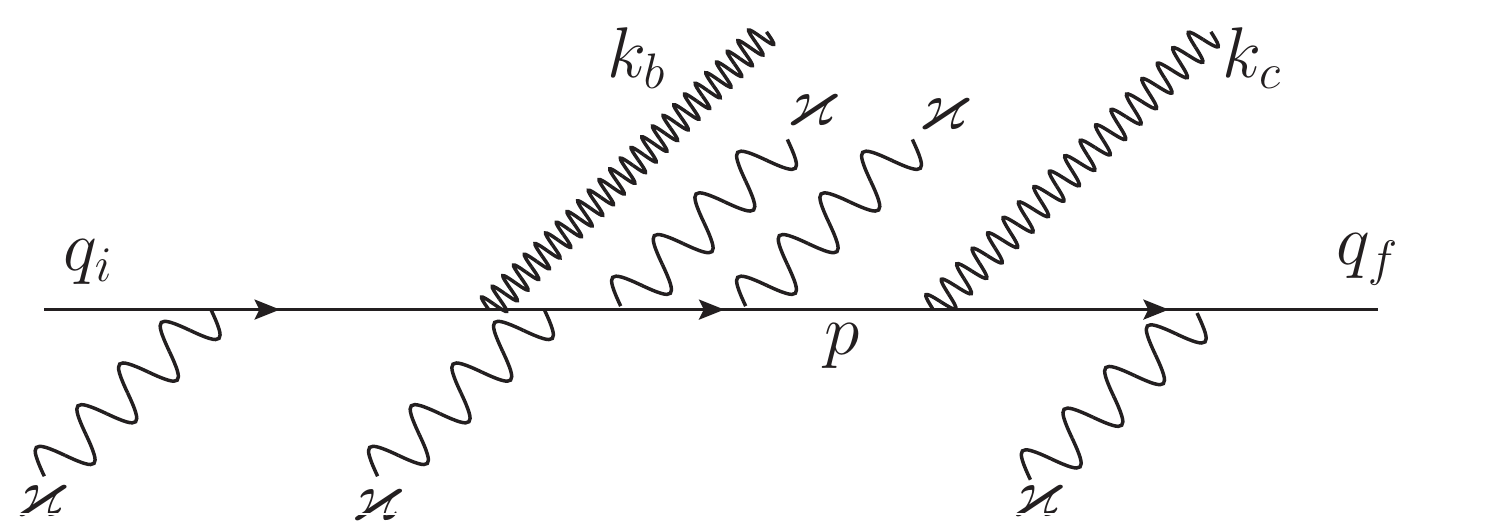}
\caption{Clarification of the index $s$. Shown above is one of the contributing
Feynman diagrams in the perturbative picture, where the laser photons are
inserted one by one. The net number of laser mode absorbed photons in this case
is $n=1$. The propagator momentum is 
$p=q_i-\kb+s\varkappa$, so that $s$ counts the net number of absorbed photons
before emitting photon $\kc$, i.e. the momentum at the position of the
label ``$p$''. For the above diagram, $s=0$.
Although $n$ must be positive for a net two-photon 
{\em emission} process, $s$ may be negative, and to get the total
amplitude for fixed $n$, one should sum all diagrams of this kind 
with $s$ ranging from $-\infty$ to
$+\infty$.
\label{s-expl}}
\end{center}
\end{figure}

%
%
\subsection{Matrix element for linear laser polarization}

The $S$-matrix element for two-photon emission from a Dirac--Volkov state
follows from standard Feynman rules, with four-vector potentials $\hat{A}_b =
A^\mu_b  \gamma_\mu = \hat{\epsilon}_b \e^{i\kb\cdot x} /\sqrt{V \, 2\omega_b}$
and $\hat{A}_c = A^\mu_c  \gamma_\mu = \hat{\epsilon}_c \e^{i\kc\cdot x}
/\sqrt{V \, 2\omega_c}$ for the two emitted photons (see also the Feynman
diagram in Fig.~\ref{fig_Feynman_diag}). For linear laser polarization, 
we get
\begin{align}
\label{twophoton_Sfi}
& S_{fi}^{\text{linear}}=S_{fi}^{(1)}+S_{fi}^{(2)}\\
& =e^2\int\ud^4x_1 \!\! \int\ud^4x_2\overline{\psi}_{q_f,r_f}(x_2)
\Big[ i\hat{A}_{c}(x_2) i G(x_2,x_1) i\hat{A}_{b}(x_1)
\nonumber\\
& \quad +i\hat{A}_{b}(x_2) iG(x_2,x_1) i \hat{A}_{c}(x_1)\Big] \,
\psi_{q_i,r_i}(x_1)
\nonumber\\
&=-i \sum_{\substack{n=1\\s=-\infty}}^\infty
\frac{(2\pi)^4e^2 m}{2V^2 \, \sqrt{\omega_c\omega_b Q_i Q_f}} \,
\delta^4(q_i-q_f+n\varkappa-\kb-\kc)
\nonumber\\
& \qquad \times \bar{u}_{r_f}(p_f) \,
\left[M_b^{s-n} \frac{\hat{{p}}_b - 
\xi^2\frac{m^2}{2\varkappa\cdot {p}_b}\hat{\varkappa} +m }
{{p}_b^2-m_\ast^2}F_b^{s}  \right.
\nonumber\\
& \qquad \qquad \left. +{M}^{s-n}_c
\frac{\hat{{p}}_c-
\xi^2\frac{m^2}{2\varkappa\cdot {p}_{c}}\hat{\varkappa} +m }%
{{p}^2_{c}-m_\ast^2}{F}_c^s\right] u_{r_i}(p_i).
\nonumber
\end{align}
Here, $G(x,y)$ denotes the laser-dressed propagator 
function~\cite{ReEb1966,Ri1985}, which can be constructed from the Volkov 
state \eqref{Volkov_lin}. The 
propagator momenta are given as
\begin{equation}
\label{def_of_tildep_if}
{p}_b=\;
q_i+s\varkappa-\kb,\qquad{p}_c=q_i+s\varkappa-\kc.
\end{equation} 
The matrix element is proportional to $V^{-2}$, since there are
one in-state and three out-states, each with a factor $\sqrt{V}$.
Here, $n$ is the {\it net} number of absorbed laser photons,
and the summation index $s$ can be understood as the 
number of laser photons absorbed up to and immediately 
before emitting the second photon (see Fig.~\ref{s-expl}
for a pictorial explanation). 

The matrix-valued functions for the transition
currents $M$ and $F$ are given as follows. 
For the first channel, we have
\begin{align}
\label{Mmatrix2}
& M_b^{s-n} =
A_0(s-n,\alpha_f-{\alpha}_b,\beta_f-{\beta}_{b}) \,
\hat{\epsilon}_{c} \nonumber\\
& \;+A_1(s-n,\alpha_f-{\alpha}_b,\beta_f-{\beta}_{b}) \,
\left(\hat{\epsilon}_{c}
\frac{e\hat{\varkappa}\hat{a}}{2\varkappa\cdot
{p}_{b}}+\frac{e\hat{a}\hat{\varkappa}}{2\varkappa\cdot p_f}
\hat{\epsilon}_{c}\right) \nonumber\\
&\;-A_2(s-n,\alpha_f-{\alpha}_b, \beta_f-{\beta}_{b})
 \frac{e^2a^2  \varkappa\cdot\epsilon_c}
{2\varkappa\cdot p_f\varkappa\cdot {p}_{b}}
\hat{\varkappa} \,
,
\end{align}
and
\begin{align}
\label{Fmatrix2}
& F_b^{s}=\;
A_0(s,\alpha_i-{\alpha}_{b},\beta_i-{\beta}_{b}) \,
\hat{\epsilon}_{b}
\nonumber\\
& \; + A_1(s,\alpha_i-{\alpha}_{b},\beta_i-{\beta}_{b}) \,
\left(\hat{\epsilon}_{b}
\frac{e\hat{\varkappa}\hat{a}}{2\varkappa\cdot p_i}+
\frac{e\hat{a}\hat{\varkappa}}{2\varkappa\cdot {p}_{b}}
\hat{\epsilon}_{b}\right)
\nonumber\\
& \; - A_2(s,\alpha_i-{\alpha}_{b},\beta_i-{\beta}_{b}) \,
\frac{e^2a^2 \varkappa\cdot \epsilon_{b}}
{2\varkappa\cdot {p}_{b} \varkappa\cdot p_i}
 \hat{\varkappa}\,.
\end{align}
For the second channel, the two currents are given as
$ M_c^{s-n}=\; M_b^{s-n}(c\leftrightarrow b) $
and $F_c^{s}= F_b^{s}(c\leftrightarrow b)$ 
under replacements of the corresponding expressions
for the first channel. The arguments entering the 
generalized Bessel functions read
\begin{equation}
\alpha_j= \; \frac{e \, a\cdot p_j}{\varkappa\cdot p_j},
\qquad
\beta_j=\frac{e^2 \, a^2}{8p_j\cdot \varkappa},
\end{equation}
with $j\in\{i,f,b,c\}$.
The spinors $u_{r_{i,f}}$ describe the spin state of the in- and outgoing
electron, respectively.  Note that $\varkappa\cdot p_{i,f}=\varkappa\cdot
q_{i,f}$, and that due to $\varkappa^2=0$, ${\alpha}_{c,b}$ and
${\beta}_{c,b}$ are independent of the summation index $s$,
although one might have initially assumed a dependence on $s$ in view of the 
presence of $p_b$ and $p_c$ in their respective defining equations.  

%
%
\subsection{Matrix element for circular laser polarization}

For the case of circular polarization of the laser, the matrix element can be derived
in a similar way to Eq.~\eqref{twophoton_Sfi}. 
The matrix element reads
\begin{align}
\label{twophoton_Sfi_circular}
& S_{fi}^{\text{circular}} = S_{fi}^{(1)c}+S_{fi}^{(2)c}\\
& =-i \sum_{\substack{n=1\\s=-\infty}}^\infty
\frac{(2\pi)^4e^2 m}{2V^2\sqrt{\omega_c\omega_b Q_i Q_f}} \,
\delta^4(q_i-q_f+n\varkappa-\kb-\kc)
\nonumber\\
& \times
\bar{u}_{r_f}(p_f)
\left[N_b^{s-n} 
\frac{\hat{{p}}_b-\xi^2\frac{m^2}{2\varkappa\cdot {p}_b}\hat{\varkappa} + m}
{{p}_b^2-m_\ast^2}G_b^{s} \right.
\nonumber\\
& \; \qquad \left. + {N}^{s-n}_c
\frac{\hat{{p}}_c-\xi^2\frac{m^2}{2\varkappa\cdot {p}_{c}}\hat{\varkappa} +m }%
{{p}^2_{c}-m_\ast^2}{G}_c^s\right] u_{r_i}(p_i) \,.
\nonumber
\end{align}
Here, as is typical for circular polarization,
the generalized Bessel functions in the formulas simplify 
to ordinary Bessel functions. The matrix-valued functions for the 
first channel read
\begin{align}
\label{Nmatrix2_circ}
&N_b^{s-n}= \;
J_{s-n}(\bar{\alpha}_{fb})\e^{i\varphi_{fb}(s-n)}
\left(\hat{\epsilon}_{c}-\frac{e^2a_1^2 \,
\varkappa\cdot \epsilon_{c}}{2\varkappa\cdot p_f \varkappa\cdot {p}_b}
\hat{\varkappa}\right) \\
&+
\Big[J_{s-n-1}(\bar{\alpha}_{fb})\e^{i\varphi_{fb}(s-n-1)}
+J_{s-n+1}(\bar{\alpha}_{fb})\e^{i\varphi_{fb}(s-n+1)}\Big]
\nonumber
\\
&\qquad\times
\frac{1}{2}\left(\hat{\epsilon}_{c}
\frac{e\hat{\varkappa}\hat{a}_1}{2\varkappa\cdot
 {p}_{b}}+\frac{e\hat{a}_1\hat{\varkappa}}{2\varkappa\cdot p_f}
\hat{\epsilon}_{c}\right)
\nonumber\\ 
&+\left[J_{s-n-1}(\bar{\alpha}_{fb})\e^{i\varphi_{fb}(s-n-1)}-
J_{s-n+1}(\bar{\alpha}_{fb})\e^{i\varphi_{fb}(s-n+1)}\right]
\nonumber
\\
&\qquad\times
\frac{1}{2i}\left(\hat{\epsilon}_{c}
\frac{e\hat{\varkappa}\hat{a}_2}{2\varkappa\cdot
 {p}_{b}}+\frac{e\hat{a}_2\hat{\varkappa}}{2\varkappa\cdot p_f}
\hat{\epsilon}_{c}\right),
\nonumber
\end{align}
and
\begin{align}
\label{Gmatrix2}
&G_b^{s}=\;
J_{-s}(\bar{\alpha}_{ib})\e^{-i\varphi_{ib}s}
\left(\hat{\epsilon}_{b}-\frac{e^2 a_1^2 \,
\varkappa\cdot \epsilon_{b}}{2\varkappa\cdot p_i \varkappa\cdot {p}_b}
\hat{\varkappa}\right)
\\
& +\left[J_{-s-1}(\bar{\alpha}_{ib})\e^{i\varphi_{ib}(-s-1)}+
J_{-s+1}(\bar{\alpha}_{ib})\e^{i\varphi_{ib}(-s+1)}\right]
\nonumber
\\
&\qquad\times
\frac{1}{2}\left(\hat{\epsilon}_{b}
\frac{e\hat{\varkappa}\hat{a}_1}{2\varkappa\cdot
 p_i}+\frac{e\hat{a}_1\hat{\varkappa}}{2\varkappa\cdot {p}_b}
\hat{\epsilon}_{b}\right)
\nonumber\\&
+\left[(J_{-s-1}(\bar{\alpha}_{ib})\e^{i\varphi_{ib}(-s-1)}-
J_{-s+1}(\bar{\alpha}_{ib})\e^{i\varphi_{ib}(-s+1)}\right]
\nonumber
\\
&\qquad\times
\frac{1}{2i}\left(\hat{\epsilon}_{b}
\frac{e\hat{\varkappa}\hat{a}_2}{2\varkappa\cdot
 p_{i}}+\frac{e\hat{a}_2\hat{\varkappa}}{2\varkappa\cdot {p}_b}
\hat{\epsilon}_{b}\right) \,.
\nonumber
\end{align}
For the second channel, we have
$N_c^{s-n}= N_b^{s-n} (b \leftrightarrow c)$
and
$G_c^{s}= G_b^{s} (b \leftrightarrow c)$.
Here,
\begin{equation}\begin{split}\label{baralpha}
&\bar{\alpha}_{fb}=\sqrt{\left(\alpha^1_{fb}\right)^2+
\left(\alpha^2_{fb}\right)^2},\quad
\bar{\alpha}_{fc}=\sqrt{\left(\alpha^1_{fc}\right)^2+
\left(\alpha^2_{fc}\right)^2},\\[2ex]
&\alpha^1_{fb}=\frac{ep_f\cdot a_1}{\varkappa\cdot p_f}-
\frac{e{p}_b\cdot a_1}{\varkappa\cdot {p}_b},\quad
\alpha^2_{fb}=\frac{ep_f\cdot a_2}{\varkappa\cdot p_f}-
\frac{e{p}_b\cdot a_2}{\varkappa\cdot {p}_b},
\end{split}\end{equation}
and similarly for $\bar{\alpha}_{ib,c}$, $\alpha^{1,2}_{ib,c}$.
The phases $\varphi$ can be expressed in terms of 
the generalized arctan function \eqref{defarctan2},
\begin{align}
\label{defpsi}
\varphi_{fb,c}=& \; 
\arctantwo{-\alpha^2_{fb,c}}{\;\alpha^1_{fb,c}},
\nonumber\\
\quad \varphi_{ib,c}=& \; 
\arctantwo{\alpha^2_{ib,c}}{\;-\alpha^1_{ib,c}}.
\end{align}
As in the linear case, the propagator momenta $p_{b,c}$ are given in
Eq.~\eqref{def_of_tildep_if}.

%
%
\subsection{Resonance conditions} 

For the whole two-photon process,
we have both momentum and energy conservation, as given by the four-dimensional
Dirac $\delta$ function in 
Eq.~\eqref{twophoton_Sfi}. The final electron is not interesting,
and therefore integrated out. Left is then one constraint from the delta function.
If this is used to fix the energy of one of the photons (we will always take photon
$\kc$ to have fixed energy), then we are free to choose the energy 
$\omega_b$ and the direction $(\theta_b,\psi_b)$ of photon $\kb$, and
the direction  $(\theta_c,\psi_c)$ of photon $\kc$. In addition, since
we are interested in polarization resolved rates, the polarization
vectors $\epsilon_b$ and $\epsilon_c$ can be chosen arbitrarily.
 The frequency  $\omega_c$ can be written as a function of the 
direction angles $\theta_b$, $\theta_c$, $\psi_b$, $\psi_c$ as follows,
\begin{align}\label{omegac}
\omega_c &= \frac{n \varkappa\cdot q_i -\kb\cdot q_i
-n\varkappa\cdot \kb}
{n\varkappa\cdot \kctilde+q_i\cdot \kctilde
-\kb\cdot \kctilde}
\nonumber\\
& \approx \frac{4n\omega E_i-\omega_b
\left[\theta_b^2
E_i+\frac{m^2}{E_i}(1+\xi^2)\right]}{\theta_c^2
E_i+\frac{m^2}{E_i}(1+\xi^2)},
\end{align}
where $\kctilde=\kc/\omega_c$. In the second line of 
Eq.~\eqref{omegac}, we have expanded the expression for 
small $\omega/m$, $\theta_b$, $\theta_c$ and ${m}/{E_i}$, and
we have assumed the conditions we are interested 
in here, i.e.~$n\frac{\omega}{m}\ll\frac{m}{E_i}
\sim \theta_b\sim\theta_c\ll1$ (this limit 
corresponds to a small total exchanged laser photon energy 
as compared to the relativistic electron energy).
Finally, the limiting term for $\theta_b =\theta_c= 0$ 
and $ \omega_b= 0$ is
\begin{equation}
\omega^{\rm max}_c 
= \frac{4 n \, \omega E_i^2}{m^2 (1 + \xi^2)}
= \frac{4 n \, \gamma^2 \omega}{(1 + \xi^2)},
\end{equation}
confirming the estimate given in Sec.~\ref{intro}.
The factor $(1+\xi^2)^{-1}$ can be interpreted simply as
arising from the increased effective mass of the 
electron in the field.

\begin{figure}[thb]
\begin{center}
\includegraphics*[width=0.4\textwidth]{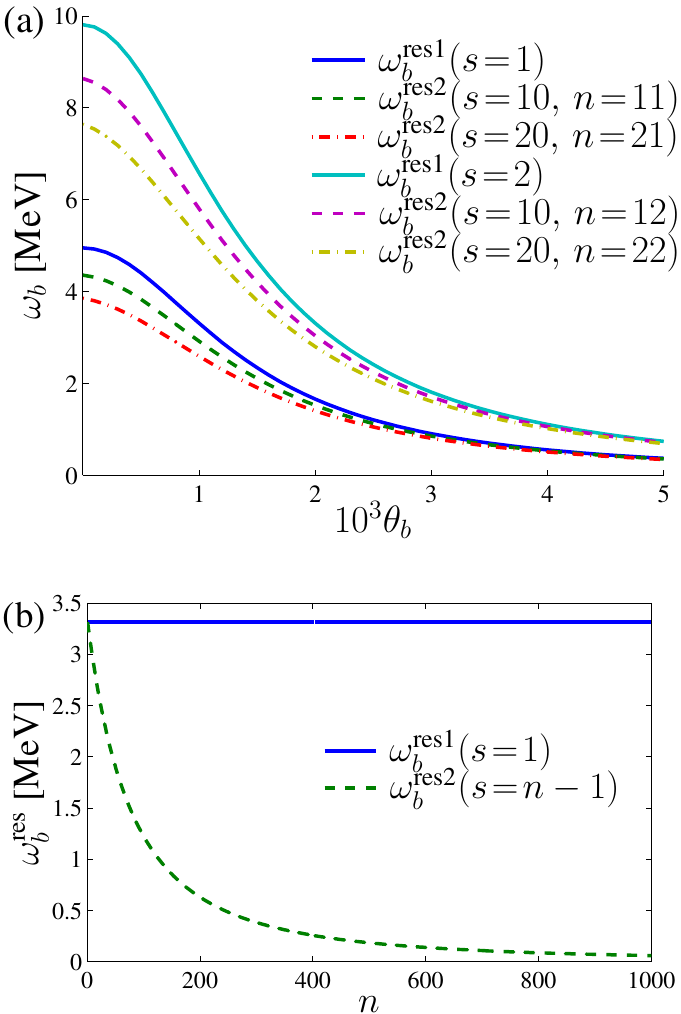}
\caption{(Color online.) Panel (a) shows $\omega_b^{{\rm res} 1}$ 
[Eq.~\eqref{first_res_cond}] and 
$\omega_b^{{\rm res} 2}$ 
[Eq.~\eqref{second_res_cond}], as a function of
$\theta_b$. Recall that we write $\boldkb= \omega_b(\sin\theta_b\sin\psi_b,
\sin\theta_b\cos\psi_b,\cos \theta_b)$, i.e.~$\theta_b$ is the angle between
$\boldkb$ and $\bm{p}_i$. The 
parameters employed are $\omega=2.5$~eV, $\xi=1$,
$E_i=10^3m$, $\psi_b=\psi_c=0$, and $\theta_c=10^{-3}$. 
In panel~(b), the resonance position of 
the first harmonic at $\theta_b=10^{-3}$ is plotted
as a function of $n$. For $\omega_b^{{\rm res} 1}$, the first harmonic
(the resonance at lowest possible $\omega_b$) means $s=1$, and 
since for $\omega_b^{{\rm res} 2}$ the value of $n-s$ tells us the order
of the resonance, we have set $s=n-1$ for this curve.  In fact, 
for large values
of $n$, the resonance $\omega_b^{{\rm res} 2}$ 
with $s=n-1$ shifts down to low photon energies, so that there will
be resonances for any photon energy $\omega_b>0$. However, these higher-order
resonances will be suppressed by a large-order Bessel function, and 
effectively, one can say that the higher-order
resonances will not contribute provided $\xi$ is not too large ($\sim 1$).
\label{respos}}
\end{center}
\end{figure}

Resonances in the Dirac--Volkov propagator \cite{Ro1985,LoJeKe2007}
occur if we have either ${p}_b^2-m_\ast^2=0$ or 
${p}_c^2-m_\ast^2=0$. Here the two-photon amplitude \eqref{twophoton_Sfi} 
splits up in a product of two single nonlinear Compton 
scattering \cite{NiRi1964,PaKaEh2002,HarHeinIld2009} amplitudes multiplied with 
a singular factor. If we solve for $\omega_b$, we find 
that the resonance conditions read
\begin{equation}
\label{first_res_cond}
\omega_b^{{\rm res}1}=\frac{s \, \varkappa\cdot q_i}%
{q_i\cdot \kbtilde +s \,\, \varkappa\cdot \kbtilde}
\approx\frac{4 s \omega E_i^2}{\theta_b^2 E_i+\frac{m^2}{E_i}(1+\xi^2)},
\end{equation}
independent of $n$ and $\kc$ (this is the usual nonlinear Compton formula 
\cite{NiRi1964}), and
a second type of resonances occurs at
\begin{equation}
\label{second_res_cond}
\begin{split}
\omega_b^{{\rm res} 2} &=
\frac{n \, \varkappa\cdot q_i - C_s(q_i \cdot \kctilde + n \, \varkappa\cdot \kctilde)}
{-C_s \, \kbtilde \cdot \kctilde +q_i\cdot \kbtilde +
n \, \varkappa\cdot \kbtilde} \,, \\[2ex]
C_s &= \frac{s \, \varkappa\cdot q_i}{q_i\cdot \kctilde +
s \, \varkappa\cdot \kctilde}.
\end{split}
\end{equation}
Equation \eqref{second_res_cond} depends on $n$, so that there is one
peak for each $n$, in principle. However, the dependence on $s$
is the decisive one for typical situations. This is natural when we recall
that $s$ is the number of photons exchanged before 
the emission of the {\em second} photon. This 
type of resonance, where the electron scatters twice inside the laser 
pulse and emits one photon at each scattering event, has been
referred to as ``plural Compton scattering'' in Ref.~\cite{BaBoetal2004}.
Figure~\ref{respos} illustrates the formulas 
\eqref{first_res_cond} and \eqref{second_res_cond}.

%
%
\subsection{\label{gauge}Via gauge invariance to the differential rate}

The matrix elements
\eqref{twophoton_Sfi} and \eqref{twophoton_Sfi_circular} are
both invariant under the gauge transformations
\begin{equation}
\epsilon_b \to \epsilon_b+\lambda_1 \, \kb,\qquad
\epsilon_c \to \epsilon_c+\lambda_2 \, \kc,
\end{equation}
where $\lambda_{1,2}$ are arbitrary constants (that may depend
on the parameters in the problem, i.e $\omega$, $q_i$ etc).
This symmetry can be used for a numerical check of
the computer code used for the evaluation, which we have performed 
in order to reassure ourselves regarding the consistency
of the calculations.
The gauge symmetry depends sensitively on the Bessel functions
and the recurrence relations satisfied by them~\cite{LoJe2009}, so that
all signs in the formulas have to be right for the symmetry to hold.
The gauge symmetry can also be used to simplify the expression, 
for example, by gauge transforming so that 
terms proportional to $\epsilon_{b,c}\cdot \varkappa$ vanish.
There is also invariance under 
the transformation $a\to a+\Lambda \varkappa$, 
$\Lambda$ constant, but since the four-vector $a$ always 
appears with a square, $a^2$, as $\hat{a}\hat{\varkappa}$, 
or in expressions like \eqref{baralpha}, this 
gauge symmetry is almost trivial and cannot be used as a 
meaningful validity check.

We now discuss how to obtain the differential two-photon rates, 
using the example of linear polarization.
The differential rate per unit time $\ud\dot W$ is obtained as 
\begin{equation}\label{rate}
\ud\dot{W}=\frac{1}{T}|S_{fi}|^2 \frac{V\ud^3 q_f}{(2\pi)^3}
\frac{ V\ud^3 \kb}{(2\pi)^3}\frac{ V\ud^3 \kc}{(2\pi)^3}.
\end{equation}
Here, $\ud^3 k_{b,c}=\omega_{b,c}^2 \ud\omega_{b,c} \ud\Omega_{b,c}$.
The squared amplitude $|S_{fi}|^2$ contains the Dirac $\delta$
of argument zero, $\left[\delta^{(4)}(0)\right]^2 =
\delta^{(4)}(q_i-q_f+n\varkappa-\kb-\kc) T V
(2\pi)^{-4}$,
so that all factors of $V$ and $T$ in~\eqref{rate} cancel, as they 
should. 
We integrate over the final electron momentum and the 
photon energy $\omega_c$ with the delta
function, and in addition we sum over the final electron spin 
(the final electron is always assumed to be unobserved), and 
average over the initial electron spin. Since in all 
examples we will present, the initial electron energy $E_i$ and 
laser intensity $\xi$ are chosen such that the quantum parameter
$\chi$ [see Eq.~\eqref{chidef}] is small, 
spin effects are marginal~\cite{Ri1985}. The
final result then reads
\begin{align}\label{final_result_diff_rate}
& \frac{\ud\dot{W}}{\ud\omega_b \ud\Omega_b \ud\Omega_c}= \;
\sum_{r_i,r_f=1}^2\sum_{n=1}^\infty
\frac{e^4 m^2\omega_b \omega_c^2}{8(2\pi)^5 Q_i
 q_f\cdot \kc}
\\ 
& \; \times
\left|\sum_{s=-\infty}^\infty\bar{u}_{r_f}(p_f)
\left[M_b^{s-n} \frac{\hat{{p}}_b+\frac{e^2a^2}%
{4k\cdot \tilde{p}_b}\hat{\varkappa} +m }
{{p}_b^2-m_\ast^2}F_b^{s} \right. \right. 
\nonumber\\
& \left. \left. +
{M}^{s-n}_c \frac{\hat{{p}}_c+\frac{e^2a^2}%
{4k\cdot {p}_{c}}\hat{\varkappa} +m }%
{{p}^2_{c}-m_\ast^2}{F}_c^s\right] u_{r_i}(p_i)
\right|^2,
\nonumber
\end{align}
evaluated with $q_f=q_i+n\varkappa-\kb-\kc$ and 
$\omega_c$ is given by the first line of Eq.~\eqref{omegac}.
Note also the factor $Q_f\omega_c /(q_f\cdot \kc)$ arising from the
delta function integration over $\omega_c$.

\begin{figure}[tb!]
\begin{center}
\includegraphics*[width=0.4\textwidth]{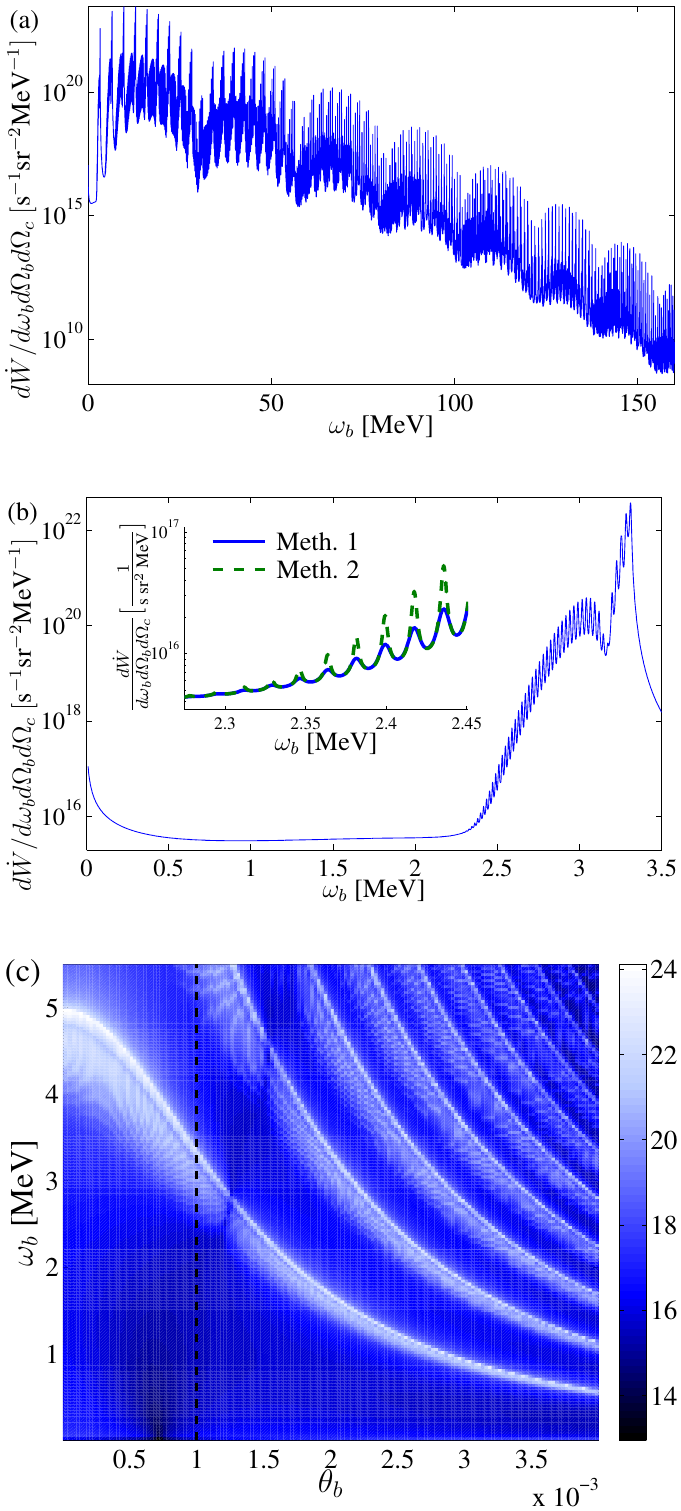}
\caption{(Color online.) We illustrate the resonances 
of the fully differential rate \eqref{final_result_diff_rate} as 
a function of $\omega_b$. 
The parameters used are $\xi=1$, $E_i=10^3m$, 
$\epsilon_b=\epsilon_b^1$,
$\epsilon_c=\epsilon_c^1$, $\theta_c=0.002$, $\psi_b=\psi_c=0$. 
In panels~(a) and~(b),
a fixed value of $\theta_b=0.001$ is used, corresponding to a cut along the
dashed line in (c). In panel (c), the color coding indicates the 
value of the  decadic logarithm 
$\log_{10}\frac{\ud\dot{W}}{\ud\omega_b \ud\Omega_b \ud\Omega_c}$,  
where the argument of the logarithm is measured in units of 
s$^{-1}$sr$^{-2}$MeV$^{-1}$. The 
regularization method employed is given by a finite laser 
pulse duration as in Eq.~\eqref{regularization2}, with
$\tau=10^4/\omega$.
In the inset of panel~(b), both methods 
\eqref{regularization1} [Meth. 1] and \eqref{regularization2} [Meth. 2]
are shown for comparison. For $\omega_b<2$ MeV, the 
application of the two methods yields
numerically indistinguishable curves.
The first Compton harmonic [the lowest bright curve in panel (c)] is 
broken at $\theta_b\approx 1.2\times10^{-3}$, which can be understood 
as the point where $\vect{a}\cdot \vect{\epsilon}_b=0$ 
in the rest frame of the electron [in this frame, and
in the gauge $\epsilon_b=(0,\vect{\epsilon}_b)$, the 
Thomson cross section is proportional to 
$|\vect{a}\cdot \vect{\epsilon}_b|^2$].
\label{spec}}
\end{center}
\end{figure}

In order to obtain a well-defined expression for the 
differential rate close to the propagator poles 
\eqref{first_res_cond}, \eqref{second_res_cond}, it 
is necessary to discuss some kind of regularization procedure.
One alternative is to include an imaginary correction
to the mass and energy of the laser-dressed electron 
\cite{BeMi1976,SchLoJeKe2007}, so that $Q_i$ and 
$m$ in the propagator denominator are replaced
according to
\begin{equation}\label{regularization1}
Q_i\to Q_i-i\frac{m\Gamma(\varkappa\cdot q_i)}{2Q_i}
,\qquad m\to m-i\frac{\Gamma(\varkappa\cdot p_{b,c})}{2}.
\end{equation}
The imaginary correction is related to the total rate for
nonlinear single Compton scattering as 
$\Gamma(\varkappa\cdot q)=\frac{q^0}{m} \dot{W}_{\textrm{Compton}}$, and
is given to a good approximation for small $\varkappa\cdot q/m^2$, 
$\xi=1$ and linear laser field polarization as 
$\Gamma(\varkappa\cdot q)=4\times 10^{-3}\varkappa\cdot q/m$
\cite{SchLoJeKe2007}. The main 
problem with this regularization scheme is that the resulting scattering 
amplitude is not strictly gauge invariant,
but the noninvariance induced by the small
regularizing imaginary parts of the energies of the virtual
states is moved to higher orders. 
We note that very similar questions concerning two-photon 
emission amplitudes for bound states have recently been 
discussed in 
\cite{UDJ2007,UDJ2008,jentschurasurzhykov2008,UDJ2009,amaroetal2009}.
As an alternative, we propose to multiply the rate with
the regularizing factor
\begin{equation}\label{regularization2}
\Phi=\prod_{s=-\infty}^{\infty}
\left[1-\e^{-\frac{\tau}{m^3}\left(p_b^2-m_\ast^2\right)^2}\right]
\left[1-\e^{-\frac{\tau}{m^3}\left(p_c^2-m_\ast^2\right)^2}\right],
\end{equation}
where $\tau$ is the pulse length of the laser field.
The rate is now proportional to $\tau$ at 
a resonance, and this way of regularizing is furthermore gauge 
invariant. Figure~\ref{spec} shows an example of the 
differential rate~\eqref{final_result_diff_rate}, evaluated 
for a specific set of parameters corresponding 
to double Compton backscattering in a relativistically 
strong laser field. For $\omega_b\gtrapprox 2$ MeV, there is
a ``forest'' of peaks at energies satisfying 
Eqs.~\eqref{first_res_cond} and \eqref{second_res_cond}.
 Note that according to Fig.~\ref{respos}~(b), the 
resonances $\omega^{\rm res2}_b(s=n-1)$ 
should actually lead to resonance peaks also
at very low photon energies, but these are suppressed by 
large-order Bessel functions and thus not visible. The 
bright curves in Fig.~\ref{spec}~(c) correspond to 
the maxima in the differential rate induced by 
single-Compton scattering, but the rate is nonvanishing 
in other areas of the $\theta_b$-$\omega_b$-plane due to the 
two-photon emission.

The object of this paper is however not to study the behavior 
of the process close to the peaks, but rather to single out 
a kinematic region where unambiguous conclusions can be drawn 
independent of the method of regularization. The kinematic region 
best suited for such investigations seems to be for photon 
energies $\omega_b$ and angles $\theta_b$ smaller than some threshold such
that the contribution from the cascade peaks are negligibly small.
Mathematically, the suppression arises due to a large-order 
generalized Bessel functions 
(or, alternatively, ordinary Bessel functions in 
the case of circular laser polarization), which beyond some 
cutoff index decays exponentially with increasing $n$ 
\cite{KoKlWi2006,LoJe2009}. In all subsequent examples
in the remaining sections of this paper, 
we will therefore restrict the photon energy $\omega_b$ 
and the polar angle $\theta_b$ to the region 
$\omega_b\leq 1$ MeV and $\theta_b\leq 0.002$. Here, 
the result is independent of the method of regularization
since we are sufficiently far away from the cascade peaks.
With increasing $\xi$, the ``safe'' region shrinks,
as the first Compton peak appears at lower energy $\omega_b$, 
see Eq.~\eqref{first_res_cond}. Already at $\xi=2$, there 
are cascade contributions at $\omega_b\le 1$ MeV, why we limit 
the laser intensity to $\xi\le 1$ in the following.

%
%
\section{Comparison to perturbative double Compton scattering}
\label{perturbative}
In the limit $\xi\to0$,
the amplitudes \eqref{twophoton_Sfi}, 
\eqref{twophoton_Sfi_circular}
reduce to the one found in \cite{MandlSkyrme1952},
where only one photon is absorbed from the laser. A discussion
of this process can be found
in standard textbooks~\cite{JauchRohr1976}, and was 
recently reexamined in~\cite{Bell2008}. The above mentioned 
references as well as other previous
works \cite{RamWang1971,Mel1972,Gould1984} were devoted to the study 
of the cross section for unpolarized initial and final photons 
(with a few exceptions, see \cite{CarPas1960,RaThu1961}). However, as noted 
in \cite{ScMai2009}, the discussion 
of photon polarization correlation necessitates an expression 
for the amplitude for arbitrary polarization of the 
final photons. 

The amplitude $S^{PDCS}$ 
for perturbative double Compton scattering (PDCS) is 
given by the sum of the Feynman diagrams shown in Fig.~\ref{FG},
and reads \cite{JauchRohr1976}
\begin{align}\label{PDCS}
S^{PDCS}=&\frac{me^3(2\pi)^4}{\sqrt{8E_i E_f\omega\omega_b\omega_c V^5}}
\delta(p_f+\kb+\kc-\varkappa-p_i)
\nonumber\\
&\times\left(\sum_{i=1}^6 N_i\right),
\end{align}
where
\begin{equation}\label{Ni}
\begin{split}
N_1&=\bar{u}(p_f) \hat{\epsilon}_c \frac{\hat{p}_f+\hatkc+m}
{(p_f+\kc)^2-m^2} \hat{\epsilon}_b \frac{\hat{p}_i+\hat{\varkappa}+m}
{(p_i+\varkappa)^2-m^2} \hat{\epsilon}u(p_i),\\
N_2&=\bar{u}(p_f) \hat{\epsilon}_c \frac{\hat{p}_f+\hatkc+m}
{(p_f+\kc)^2-m^2} \hat{\epsilon} \frac{\hat{p}_i-\hatkb+m}
{(p_i-\kb)^2-m^2} \hat{\epsilon}_b u(p_i),\\
N_3&=\bar{u}(p_f) \hat{\epsilon} \frac{\hat{p}_f-\hat{\varkappa}+m}
{(p_f-\varkappa)^2-m^2} \hat{\epsilon}_c \frac{\hat{p}_i-\hatkb+m}
{(p_i-\kb)^2-m^2} \hat{\epsilon}_bu(p_i) \,,
\end{split}\end{equation}
$N_4=N_1(b\leftrightarrow c)$, $N_5=N_2(b\leftrightarrow c)$
and $N_6=N_3(b\leftrightarrow c)$ in a self-explanatory notation.
Here, $\epsilon$ represents the polarization vector of the laser field.
The rate is then
\begin{align}\label{rate_PDCS_0}
& \ud\dot{W} =J\left|S^{PDCS}\right|^2\frac{1}{T}
V\frac{V\ud^3 \kb}{(2\pi)^3}
\frac{V\ud^3 \kc}{(2\pi)^3}\frac{V\ud^3 p_f}{(2\pi)^3}
\\
&=\frac{\xi^2m^4e^4}{8E_i E_f\omega_b\omega_c (2\pi)^5}
\left|\sum_{i=1}^6 N_i \right|^2 \delta(p_f+\kb+\kc-p_i-\varkappa)
\ud^3 \kb
\ud^3 \kc \ud^3 p_f
,
\nonumber
\end{align}
where $J=\frac{\omega m^2 \xi^2}{e^2}$ is the photon 
flux.
Integrating over $\ud^3 p_f$ and $\ud\omega_c$, and
summing and averaging over the electron spin,
we obtain
\begin{equation}\label{rate_PDCS}
\frac{\ud\dot{W}^{PDCS}}{\ud\omega_b\ud\Omega_b\ud\Omega_c}
=\xi^2\frac{m^4e^4\omega_b\omega_c^2}{16 (2\pi)^5 E_i p_f\cdot \kc}
\sum_{r_i,r_f=1}^2\left|\sum_{j=1}^6 N_j \right|^2.
\end{equation}
The dependence on $\xi$ is thus given by the prefactor $\xi^2$,
and we may thus refer to this process as the ``linear''
process because the rate is proportional
to the laser intensity.

\begin{figure*}[thb]
\begin{center}
\includegraphics*[width=0.7\linewidth]{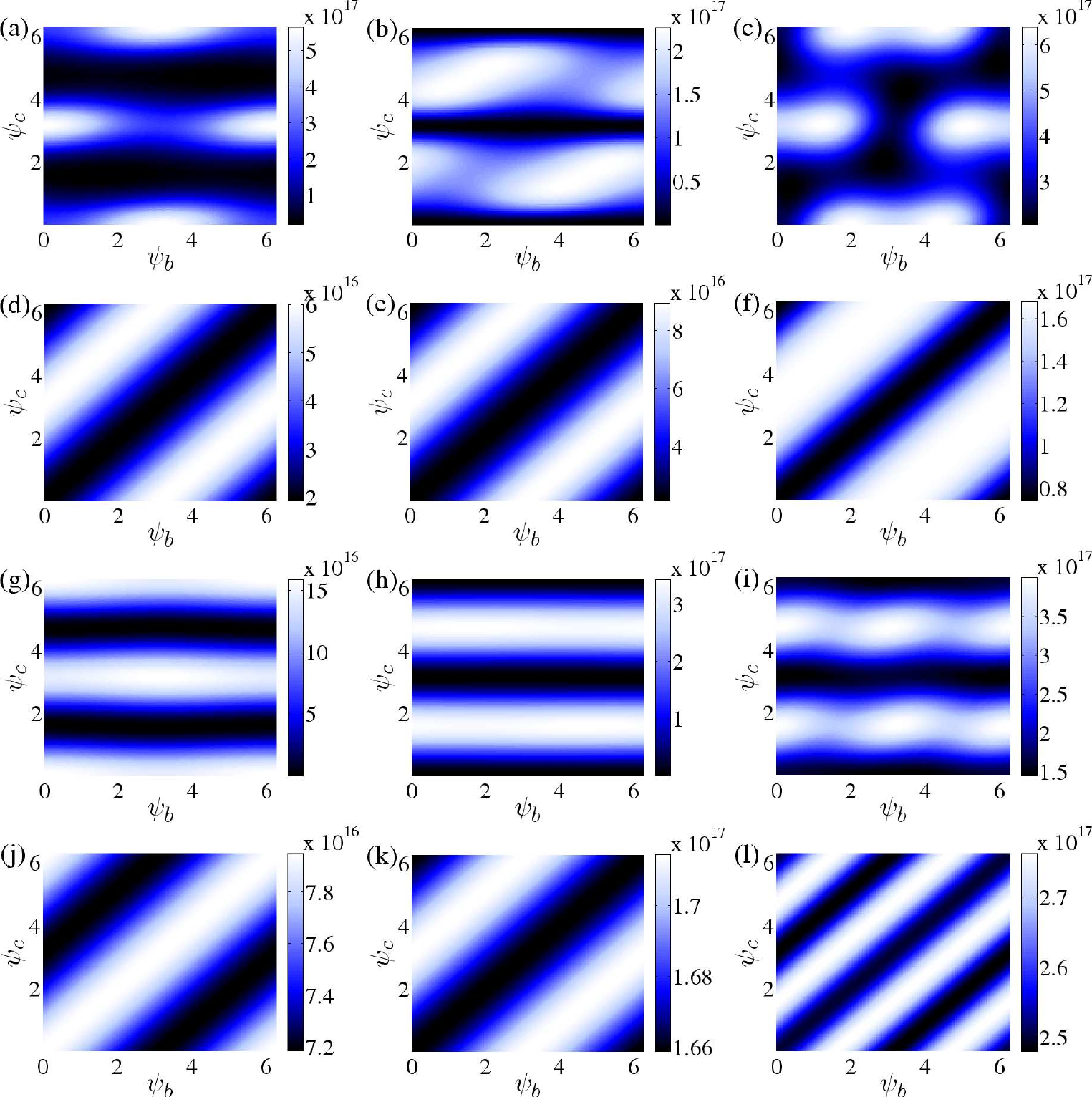}
\caption{(Color online.) Contour plot of the differential 
rate for different final photon polarizations. The color 
coding indicates the value of the differential rate
$\frac{\ud\dot{W}}{\ud\omega_b \ud\Omega_b \ud\Omega_c}$, on 
linear scale, in units of 
s$^{-1}$sr$^{-2}$MeV$^{-1}$.
The upper two rows [(a)--(f)] display the result of the nonperturbative
formula, and the lower two rows [(g)--(l)] show the perturbative results.
The parameters used are  $\xi=1$, $\omega=2.5$ eV, $E_i=10^3m$, 
$\omega_b=1$ MeV, 
and $\theta_b=2\theta_c=10^{-3}$. In this case, 
$\omega_c$ is fixed by the scattering geometry but still 
depends on the number $n$ of exchanged photons; we
present the differential rate summed over all $n$ and thus suppose 
that the energy $\omega_c$ is unobserved.
For the polarizations, we have
linear laser polarization in (a)--(c) and (g)--(i) (first and third row), 
and circular laser polarization in (d)--(f) and (j)--(l) (second and fourth row).
In (a), (d), (g), and (j) (first column) we have $\epsilon_b=\epsilon_b^1$, 
$\epsilon_c=\epsilon_c^1$, where we recall that the 
polarization vectors are defined in Eq.~\eqref{polvecs}.
In (b), (e), (h), and (k) (middle column) we have $\epsilon_b=\epsilon_b^1$, 
$\epsilon_c=\epsilon_c^2$, and 
panels (c), (f), (i) and (l) (right column) show the differential rate 
summed over all possible polarizations of the final photons.
 \label{psibpsic}}
\end{center}
 \end{figure*}

\begin{figure*}[thb]
\begin{center}
\includegraphics*[width=0.7\textwidth]{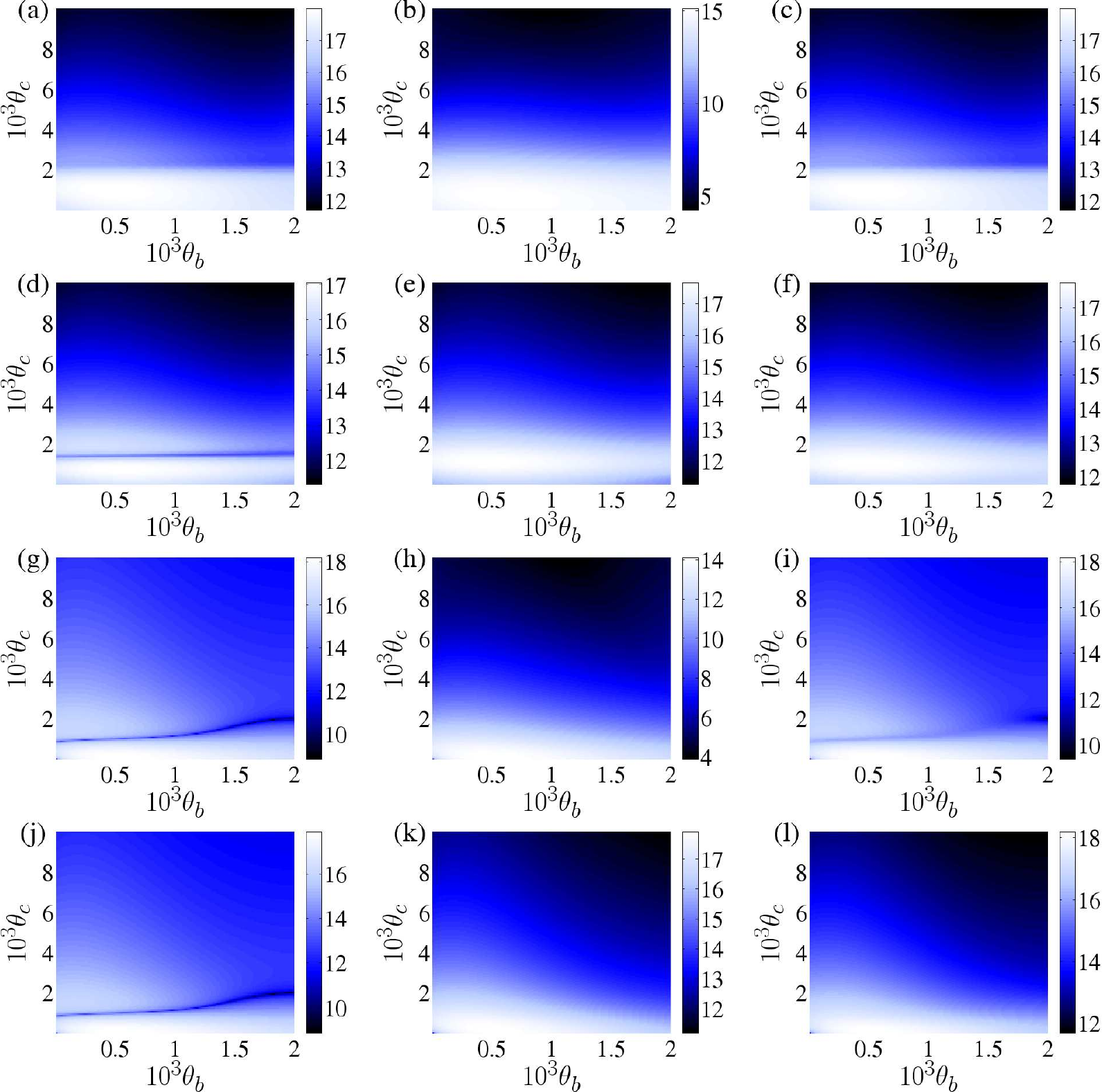}
\caption{(Color online.) Contour plot of the differential 
rate as a function of the polar angles $\theta_b$ and 
$\theta_c$. The color 
coding indicates the value of the decadic logarithm 
$\log_{10}\frac{d\dot{W}}{d\omega_b d\Omega_b d\Omega_c}$, with 
the differential rate given in units of 
s$^{-1}$sr$^{-2}$MeV$^{-1}$.
As in Fig.~\ref{psibpsic}, the upper two rows [(a)--(f)] show nonperturbative, 
and the lower two rows [(g)--(l)] perturbative results, respectively.
The parameters used are  $\xi=1$, $\omega=2.5$ eV, $E_i=10^3m$, 
$\omega_b=1$ MeV, 
and $\psi_b=0$, $\psi_c=\pi$. For the polarizations, we have
linear laser polarization in (a)--(c) and (g)--(i) (first and third row), 
and circular laser polarization in (d)--(e) and (j)--(l) (second and fourth row).
In (a), (d), (g), and (j) (left column) we have $\epsilon_b=\epsilon_b^1$, 
$\epsilon_c=\epsilon_c^1$, 
In (b), (e), (h), and (k) (middle column) we have $\epsilon_b=\epsilon_b^1$, 
$\epsilon_c=\epsilon_c^2$, and 
panels (c), (f), (i) and (l) (right column) show the differential rate 
summed over the final photon polarizations. Note that from 
Fig.~\ref{polv}, we have 
$\epsilon^1_b\cdot\epsilon^1_c\approx1$ and 
$\epsilon^1_b\cdot\epsilon^2_c\approx0$ here.
 \label{thetabthetac}}
\end{center}
\end{figure*}

In Figs.~\ref{psibpsic} and \ref{thetabthetac}, we show a comparison of the 
predictions of the nonperturbative formulas 
[Eq.~\eqref{twophoton_Sfi} and Eq.~\eqref{twophoton_Sfi_circular}]
and the perturbative formula [\eqref{PDCS}] for the differential rate, 
for both circular and linear polarization of the laser field. Note that
to compare with the circular polarization, one should put 
$\epsilon=(0,1,i,0)/\sqrt{2}$ in Eq.~\eqref{Ni}. The figures show 
that the photons may be produced in any polarization state, 
although parallel polarization of the two emitted
photons $\epsilon_b\cdot\epsilon_c \approx 1$ is the 
dominant channel for linear laser polarization. Due to the rotational symmetry,
circular laser polarization results in similar differential rates for 
both parallel and perpendicular polarization of the emitted high-energy photons
[see Figs.~\ref{psibpsic}(d)--(f) and (j)--(l)].
 For small laser intensity, the 
plane defined by the polarization and propagation axes of the 
laser characterizes the emission pattern of the accelerated charge,
so that alignment of the two emitted photons as shown in the 
third row of Fig.~\ref{psibpsic} can be expected. 
Our numerical results show that this intuitive
picture is still valid in the relativistic, nonperturbative 
laser interaction regime, although some details change:
e.g., the emission of photons with antiparallel polarization
vectors is favored over the parallel case, as is 
evident from the first panel in the upper row of Fig.~\ref{psibpsic}.
We here recall that the direction of the polarization 
basis vectors $\epsilon_{b,c}^{1,2}$ depends on the angles
 $\psi_{b,c}$, see Fig.~\ref{polv}. Figure~\ref{thetabthetac} 
confirms this picture, here the difference of the differential rate
between the parallel case ($\epsilon_b\cdot \epsilon_c\approx 1$) 
and the perpendicular case ($\epsilon_b\cdot \epsilon_c\approx 0$) 
amounts to
several orders of magnitude. 

\begin{figure}[tbh]
\begin{center}
\includegraphics*[width=0.8\columnwidth]{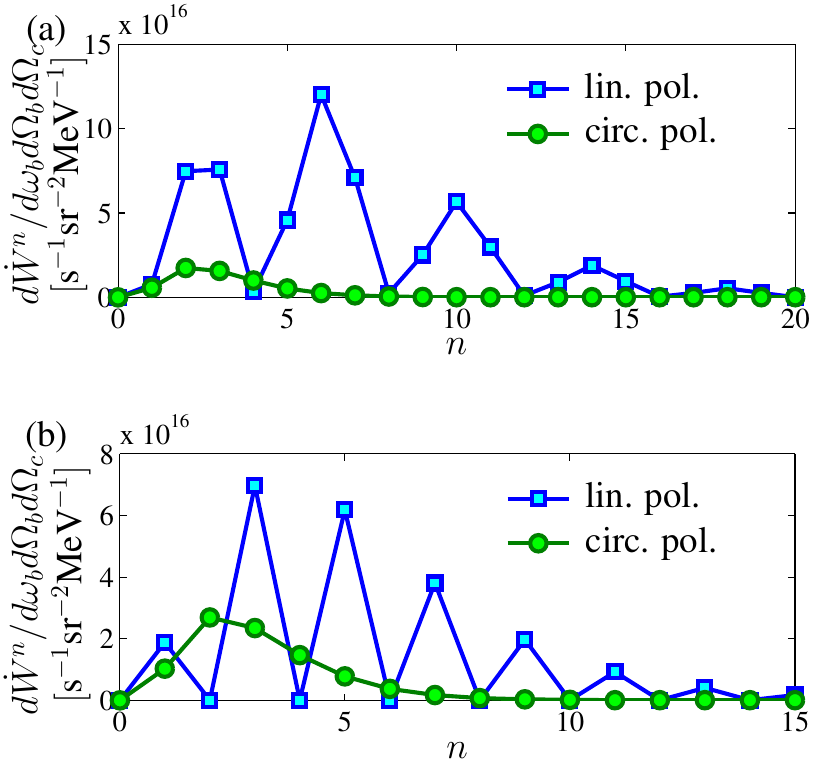}
\caption{(Color online.) Demonstration of the 
multiphoton character of the pair creation rate, for 
linear and circular polarization of the laser field.
In (a), $\psi_b=0$, $\psi_c=\pi$,  
and the final polarization is fixed to $\epsilon_b=\epsilon_b^1$,
$\epsilon_c=\epsilon_c^1$, while in (b) we have 
$\psi_b=\pi/2$, $\psi_c=3\pi/2$,  
and $\epsilon_b=\epsilon_b^1$,
$\epsilon_c=\epsilon_c^2$. Otherwise the parameters are the same as 
in Fig.~\ref{psibpsic}.
\label{Nplot}}
\end{center}
\end{figure}

In order to demonstrate the contribution from the different 
photon orders $n$, we
refer to Fig.~\ref{Nplot}. Here the dependence on the photon order $n$ 
is shown,
if we define
\begin{equation}
\frac{\ud\dot{W}}{\ud\omega_b \ud\Omega_b \ud\Omega_c}=
\sum_{n=1}^{\infty}
\frac{\ud\dot{W}^n}{\ud\omega_b \ud\Omega_b \ud\Omega_c}.
\end{equation}
As can be seen from Fig.~\ref{Nplot}, typically up to 20 photons contribute to
the differential rate. Equation~\eqref{omegac} for $n\leq20$ then yields
$\omega_c\leq 70$ MeV, which implies that even though the ``first'' photon has
modest energy $\omega_b=1$ MeV, the energy of the ``second'' photon is much
larger.  The difference between the smooth curve in the circular case and the
sawtooth shape of the linear curve can be traced back to the behavior of the
generalized Bessel function and the usual Bessel function, constituting the
amplitudes \eqref{twophoton_Sfi} and \eqref{twophoton_Sfi_circular}.  For
example, for the parameters shown in Fig.~\ref{Nplot}(b), one can show that the
dominant contribution to the matrix element for linear polarization is
roughly proportional to the generalized Bessel function  $A_1(n,0,\beta)$,
which vanishes for even $n$ \cite{LoJe2009}. However, if the polarization
vectors are summed over, then the case of even $n$ contributes, and the curve
smoothens out. Similar selection rules for the emitted harmonics occur also for
the nonlinear single Compton scattering process \cite{Ri1985}.  On the
contrary, the circular polarization curve is smooth due to the rotational
symmetry.

 
\begin{figure}[tbh!]
\begin{center}
\includegraphics*[width=0.8\columnwidth]{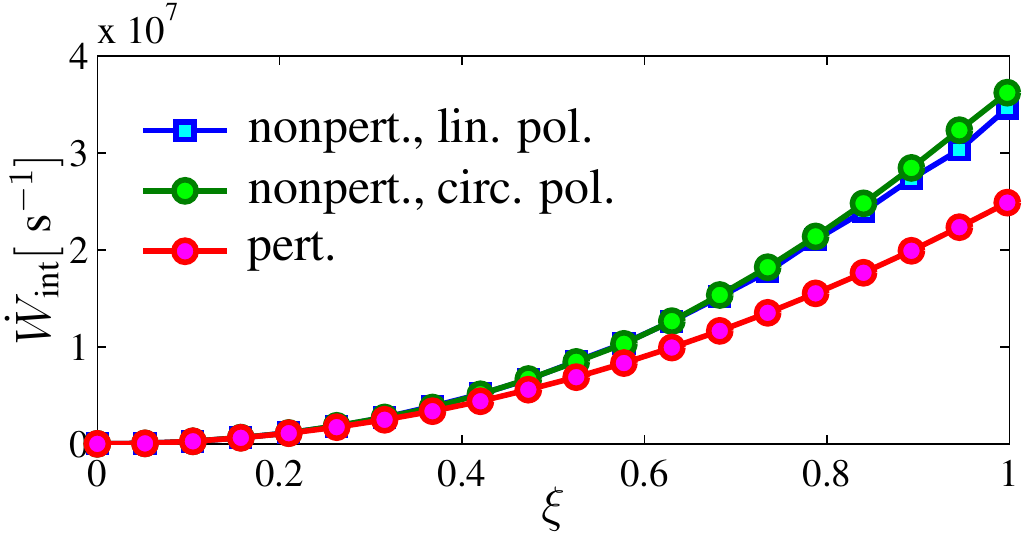}
\caption{(Color online.) The integrated rate~\eqref{Integrated_Rate},
defined in Eq.~\eqref{Integrated_Rate} for $\omega=2.5$ eV and $E_i=10^3 m$. 
The figure compares the nonperturbative formula for linear laser polarization
[Eq.~\eqref{final_result_diff_rate}], for circular laser polarization 
[Eq.~\eqref{twophoton_Sfi_circular}], and the perturbative 
expression [Eq.~\eqref{rate_PDCS}]. 
\label{Wtot}}
\end{center}
\end{figure}

To conclude this section, we investigate if the integrated rate
differ in the perturbative and nonperturbative case. In Fig.~\ref{Wtot},
we show, as a function of $\xi$, the quantity
\begin{equation}\label{Integrated_Rate}
\begin{split}
\dot{W}_{\textrm{int}}={}&\sum_{\lambda_b,\lambda_c}
\int\limits_0^{2\pi}\ud\psi_b 
\int\limits_0^{2\pi}\ud\psi_c
\int\limits_0^{\theta_{b,\textrm{max}}} \sin{\theta_b}\ud\theta_b
\\&\times
\int\limits_0^{\theta_{c,\textrm{max}}} \sin{\theta_c}\ud\theta_c
\int\limits_{\omega_{b,\textrm{min}}}^{\omega_{b,\textrm{max}}}
\ud\omega_b \,
\frac{\ud\dot{W}}{\ud\omega_b \ud\Omega_b \ud\Omega_c},
\end{split}
\end{equation}
with $\theta_{b,\textrm{max}}=1.5\times 10^{-3}$, 
$\theta_{c,\textrm{max}}=2.5\times 10^{-3}$, 
$\omega_{b,\textrm{min}}=10^{-3}$ MeV, and 
$\omega_{b,\textrm{max}}=1$ MeV.
This restriction is identical to the one in \cite{LoJe2009_2}. By 
restricting the final phase space, one can ensure that contributions from
the single Compton scattering cascade are negligible, as discussed in 
Sec. \ref{gauge}. Figure~\ref{Wtot} reveals that the integrated
rate is slightly larger than one would expect from the perturbative
formula, and also that circular and linear polarization of the 
laser gives almost identical results, despite their different
angular characteristics (see Figs.~\ref{psibpsic},~\ref{thetabthetac}).
Another remark is that for the integrated rate, the perturbative
formula gives identical results regardless of laser polarization
because interference terms in the expanded perturbative 
rate vanish after the integration. In the interval 
considered ($\xi<1$), the relative difference 
 of the 
integrated nonperturbative and the integrated perturbative rate 
can be approximately fitted to a power law as
$\dot{W}_{\textrm{int}}^{\textrm{nonpert}}-
\dot{W}_{\textrm{int}}^{\textrm{pert}}\propto \xi^\eta$, with
$\eta\approx 2.7$ for linear and $\eta\approx 3$ for circular 
laser polarization, respectively. 

%
%
%
\section{Angular correlation and entanglement}
\label{angular}

We now turn our attention to the important 
questions regarding the quantum mechanical 
correlation, i.e.~entanglement, of the two final photons.
The theory we apply in this section have been previously 
used extensively
to characterize the final-state correlation in bound
states transitions~\cite{radtkeetal2006,radtke2008,Radke2008_2,%
MaiSuretal2009}.
The idea is to use the information contained in the 
matrix elements \eqref{twophoton_Sfi}, \eqref{twophoton_Sfi_circular},
to obtain an expression for the {\it density matrix} $\rho_f$ of
the polarizations of the final system ``electron+two photons''. 
Given an expression
for $\rho_f$, it is then straightforward to calculate the 
concurrence \cite{Wootters1998}, which is a measurement of how much the
two photons are entangled.
The starting point is the initial density matrix \cite{Blum1996}
\begin{equation}
\rho_i=\sum_{r_i=1}^2|r_i,0,0\rangle\langle r_i,0,0|,
\end{equation}
where $r_i$ is the spin of the initial electron and 
the zeros denote the absence of photons (other than laser photons of course)
in the initial state. The initial electron is thus assumed to be 
unpolarized. Note also that all dependencies on energies and angles etc 
of the state 
vectors are not written out. Next, due to the interaction $R$, the density matrix
$\rho_i$ evolves into the final state density matrix $\rho_f$,
\begin{equation}
\rho_f=R\rho_i R^\dagger=\sum_{r_i=1}^2
R|r_i,0,0\rangle\langle r_i,0,0|R^\dagger.
\end{equation}
The matrix elements of $\rho_f$ are thus given by
\begin{align}
& \langle r_f,\lambda_b,\lambda_c |\rho_f |r'_f,\lambda'_b,\lambda'_c\rangle
\nonumber\\
& = \sum_{r_i=1}^2
\langle r_f,\lambda_b,\lambda_c |R|r_i,0,0\rangle\langle r_i,0,0|R^\dagger
|r'_f,\lambda'_b,\lambda'_c\rangle,
\end{align}
where $\lambda_{b,c}$, $\lambda'_{b,c}\in\{1,2\}$ denotes the polarization
components of the emitted photons in either Cartesian or circular basis.
If the final electron is unobserved, we should trace out $r_f$:
\begin{align}
& \langle \lambda_b,\lambda_c |\rho_f |\lambda'_b,\lambda'_c\rangle
\nonumber\\
& = \sum_{r_i,r_f=1}^{2}\langle r_f,\lambda_b,\lambda_c |R|r_i,0,0\rangle \,
\langle r_i,0,0|R^\dagger
|r_f,\lambda'_b,\lambda'_c\rangle.
\end{align}
If we now identify
\begin{align}
\langle r_f,\lambda_b,\lambda_c |R|r_i,0,0\rangle
=& \; \sqrt{N}S_{fi}(r_i,r_f,\lambda_b,\lambda_c),
\nonumber\\
\langle r_i,0,0|R^\dagger |r_f,\lambda'_b,\lambda'_c\rangle 
=& \; \sqrt{N}S_{fi}^\ast(r_i,r_f,\lambda'_b,\lambda'_c),
\end{align}
where $N$ is a normalization constant, 
and we use the explicit basis
\begin{equation}\begin{split}
|1\;1\rangle&=\binom{1}{0}\otimes\binom{1}{0}=
\left(\begin{array}{c}1\\0\\0\\0\end{array}\right), \\
|1\;2\rangle&=\binom{1}{0}\otimes\binom{0}{1}=
\left(\begin{array}{c}0\\1\\0\\0\end{array}\right),\\
|2\;1\rangle&=\binom{0}{1}\otimes\binom{1}{0}=
\left(\begin{array}{c}0\\0\\1\\0\end{array}\right), \\
|2\;2\rangle&=\binom{0}{1}\otimes\binom{0}{1}=
\left(\begin{array}{c}0\\0\\0\\1\end{array}\right),
\end{split}\end{equation}
for the polarization state of the final photons,
then the expression for the final density $4\times4$ matrix
reads
\begin{equation}
\rho_f=
\sum_{r_i,r_f=1}^2
N\left(\begin{array}{cccc}
S_{11}S_{11}^\ast & S_{11}S_{12}^\ast & S_{11}S_{21}^\ast & S_{11}S_{22}^\ast\\
S_{12}S_{11}^\ast & S_{12}S_{12}^\ast & S_{12}S_{21}^\ast & S_{12}S_{22}^\ast\\
S_{21}S_{11}^\ast & S_{21}S_{12}^\ast & S_{21}S_{21}^\ast & S_{21}S_{22}^\ast\\
S_{22}S_{11}^\ast & S_{22}S_{12}^\ast & S_{22}S_{21}^\ast & S_{22}S_{22}^\ast
\end{array}\right),
\end{equation}
where
\begin{equation}
S_{\lambda_b\lambda_c}=S_{fi}(r_i,r_f,\lambda_b,\lambda_c).
\end{equation}
The normalization constant $N$ can be found by requiring
\begin{equation}
1=\Tr \rho_f=\sum_{r_i,r_f}N\left(|S_{11}|^2+|S_{12}|^2+|S_{21}|^2+|S_{22}|^2\right).
\end{equation}
According to~\cite{Wootters1998}, the concurrence $C(\rho_f)$
is now given by
\begin{equation}\label{Concurrence}
C(\rho_f)=\textrm{max}(0,\sqrt{\zeta_1}-\sqrt{\zeta_2}-\sqrt{\zeta_3}
-\sqrt{\zeta_4}),
\end{equation}
where the $\zeta_j$'s are the eigenvalues, in descending order, of the matrix
\begin{equation}
Q=\rho_f \left(\sigma^2\otimes\sigma^2\right) 
\rho_f^\ast \left(\sigma^2\otimes\sigma^2\right),
\end{equation}
where $\sigma^2$ is the second Pauli spin matrix.
The eigenvalues $\zeta_j$ are real and positive. The concurrence 
as defined in Eq.~\eqref{Concurrence} is gauge invariant, and
does not depend on the basis used for the polarization 
vectors of the photons, i.e.~either the Cartesian basis, 
Eq.~\eqref{polvecs}, or the helicity basis, Eq.~\eqref{polvecs_hel}, 
can be used.
An explicit expression for $\left(\sigma^2\otimes\sigma^2\right)$ as a 
$4\times 4$ matrix is given by
\begin{equation}
\left(\sigma^2\otimes\sigma^2\right)=
\left(\begin{array}{cc}
0 & -i\\
i & 0 \end{array}\right)
\otimes\left(\begin{array}{cc}
0 & -i\\
i & 0 \end{array}\right)=
\left(\begin{array}{cccc} 
0 & 0 & 0 & -1\\
0 & 0 & 1 & 0\\
0 & 1 & 0 & 0\\
-1& 0 & 0 & 0 \end{array}\right).
\end{equation}
The matrix $\left(\sigma^2\otimes\sigma^2\right)$ is a kind of spin-flip operator
for qubits,
we have
\begin{equation}\begin{split}
\left(\sigma^2\otimes\sigma^2\right)|1\;1\rangle=-|2\;2\rangle,&\quad
\left(\sigma^2\otimes\sigma^2\right)|2\;2\rangle=-|1\;1\rangle,\\
\left(\sigma^2\otimes\sigma^2\right)|1\;2\rangle=|2\;1\rangle,&\quad
\left(\sigma^2\otimes\sigma^2\right)|2\;1\rangle=|1\;2\rangle.
\end{split}
\end{equation}
This means that a maximally entangled pure state is an eigenstate of
$\left(\sigma^2\otimes\sigma^2\right)$:
\begin{equation}
\left(\sigma^2\otimes\sigma^2\right)\left(|1\;2\rangle-|2\;1\rangle
\right)=-|1\;2\rangle+|2\;1\rangle,
\end{equation}
and consequently has unity concurrence. 

\begin{figure}[tbh]
\begin{center}
\includegraphics*[width=0.45\textwidth]{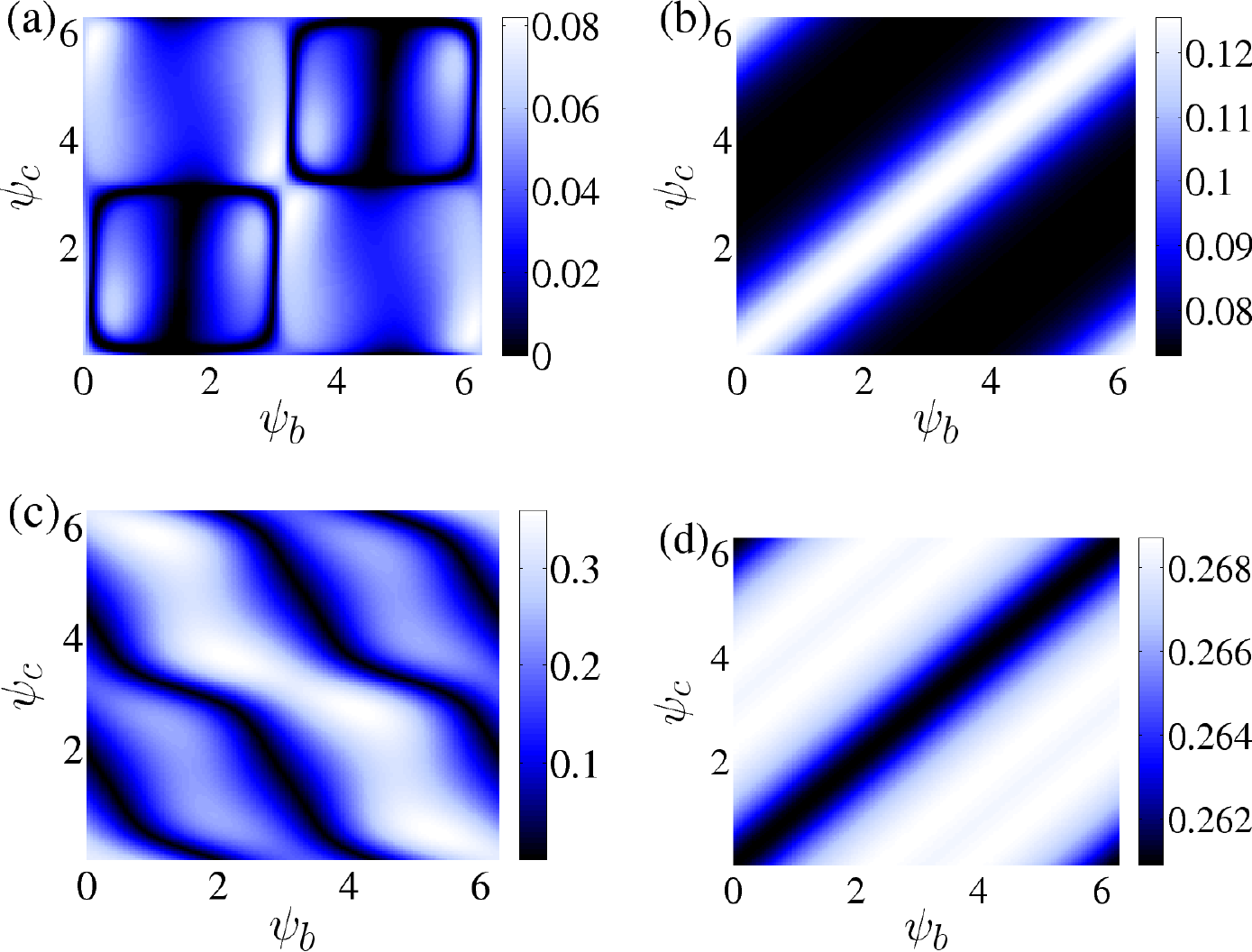}
\caption{(Color online.) Contour plot of concurrence
as a function of the azimuth angles $\psi_b$ and 
$\psi_c$. The color 
coding indicates the value of the concurrence $C(\rho_f)$
 [see Eq.~\eqref{Concurrence}]. Panel (a) shows 
linear laser polarization, and (b) circular laser 
polarization, both calculated with the nonperturbative
expressions for the amplitude. The results of 
the perturbative formula are displayed in panel (c)
[linear laser polarization] and (d) [circular laser polarization].
The values of $E_i$, $\omega$, $\xi$ and $\theta_{b,c}$ are the same 
as in Fig.~\ref{psibpsic}.
 \label{C_psibpsic}}
\end{center}
\end{figure}

\begin{figure}[tbh]
\begin{center}
\includegraphics*[width=0.45\textwidth]{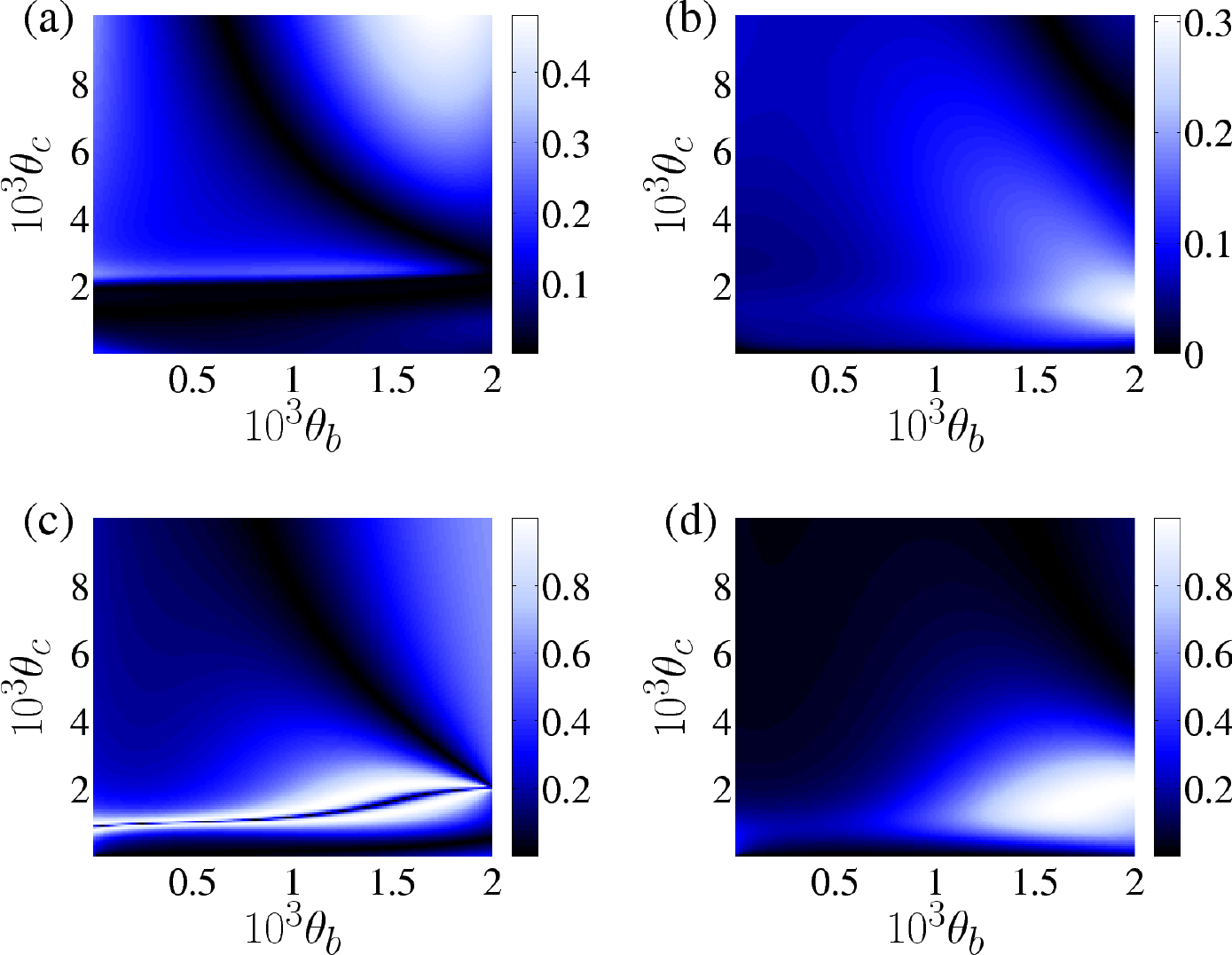}
\caption{(Color online.) Contour plot of concurrence
as a function of the polar angles $\theta_b$ and 
$\theta_c$. The color 
coding indicates the value of the concurrence $C(\rho_f)$
 [see Eq.~\eqref{Concurrence}]. Panel (a): linear laser polarization, 
nonperturbative, (b): circular laser polarization, nonperturbative, 
(c): linear laser polarization, 
perturbative, (d): circular laser polarization, perturbative.
Laser parameters as 
well as the initial electron energy $E_i$ and the 
photon azimuth angles $\psi_{b,c}$ are the same
as in Fig.~\ref{thetabthetac}. 
 \label{C_thetabthetac}}
\end{center}
\end{figure}

We now provide some examples and compare the concurrence \eqref{Concurrence}
for different laser polarizations and furthermore show that the nonperturbative
treatment is indispensable to correctly predict the degree of correlation.
Figure~\ref{C_psibpsic} shows the concurrence as a function of the azimuth
angles $\psi_b$, $\psi_c$. This figure should be compared to
Fig.~\ref{psibpsic}. In Fig.~\ref{C_thetabthetac}, we show instead the
dependence on the polar angles $\theta_{b,c}$, which should be compared with
the corresponding Fig.~\ref{thetabthetac} for the differential rate. 

We remark that
to be able to measure the concurrence, it is desirable to find angular
regions where high concurrence and high differential rate overlap. This seems
to be possible, at least in some cases: e.g., one may compare
Fig.~\ref{C_thetabthetac}(c) with Fig.~\ref{thetabthetac}(l).  Moreover, the
general trend is that a strong laser field diminishes the concurrence.
Therefore, if high entanglement is sought, it is advisable to employ a
perturbative laser beam ($\xi<1$), although the nonperturbative dependence of
the concurrence as a function of $\xi$ would be highly interesting to measure.
Similar conclusions as those above follow from the previous investigation
\cite{LoJe2009_2}.  A final remark is that linear and circular laser
polarization are seen to lead to similar peak values of the concurrence.

%
%
\section{Conclusions}
\label{conclu}

Our treatment of double Compton scattering in intense laser fields
in based on the canonical formalism of Furry-picture 
quantum electrodynamics (QED), where a strong external field
(in this case, the oscillatory laser field) is incorporated 
into the fermion propagators. The oscillatory nature 
of the laser field necessitates the expansion of all
initial and final fermion states into plane waves,
thereby giving rise to generalized (linear polarization)
and ordinary (circular polarization) Bessel functions.
The formalism, as outlined in Sec.~\ref{formulation},
leads to a consistent formulation of the nonperturbative
double Compton scattering for an arbitrary intensity of the 
laser field, and with full account of all relativistic and 
spin-dependent effects on the electron lines.
In particular, a suitable generalization of the formalism 
outlined here would apply to three-photon events,
which can be described by a third-order amplitude 
in QED.

In addition to a consistent formulation
of the polarization resolved production rates, 
differential in the photon emission angles and energy,  
for two-photon transitions of Dirac--Volkov states 
in intense laser fields, we numerically show that 
only a fully relativistic formalism, nonperturbative in the 
laser field strength, can possibly yield experimentally 
verifiable, consistent predictions.
This is not surprising because the differential
rates depend crucially on details of the 
emission process, which in turn is highly dependent
on the properties of the propagators near the resonances.
Indeed, it is possible to identify those angular
and photon energy regions where only the two-photon
amplitude, not the single-Compton resonances, give 
appreciable contributions to the photon emission,
and it is thus possible to observe entangled high-energy 
photons in coincidence without having to worry about 
background from resonant cascade emission by
single-photon transitions.

The necessity of the nonperturbative formalism is demonstrated in 
Sec.~\ref{perturbative}.
The perturbative (in the laser field interaction)
double Compton scattering cannot give reliable 
predictions if the nonlinear intensity parameter $\xi$
approaches unity. We stress that a value 
$\xi\approx 1$, corresponding to laser intensities
of the order of $10^{18}$ W/cm$^2$ for optical 
lasers, is routinely available today in 
many laboratories worldwide. In Figs.~\ref{psibpsic}
and~\ref{thetabthetac}, we show that depending on polarization
and the observation solid angle, even order-of-magnitude
differences can exist between the rates evaluated
with the nonperturbative and the perturbative formulas.
However, if angles and photon energies
are integrated over, the results are similar, as shown in 
Fig.~\ref{Wtot}, although the difference grows nonlinearly 
with $\xi$, illustrating the importance of higher orders.

The polarization entanglement is interesting but needs to be 
quantified. Therefore, we discuss, in Sec.~\ref{angular},
the concurrence as a gauge-independent measure of the 
photon entanglement. Our results (see Figs.~\ref{C_psibpsic}
and~\ref{C_thetabthetac}) indicate that close to maximally 
entangled (unity concurrence) photon pairs may be produced, but
only in certain angular regions. Furthermore, the 
degree of entanglement changes strongly with the laser
field intensity. An experimental verification of the entanglement
would yield a test for this fundamental quantum phenomenon in 
a high-energy domain where it is otherwise difficult to 
generate entangled quanta.

Finally, we remark that the two-photon emission is not 
a ``rare'' or ``unusual'' physical process but a simple generalization of the basic
physical phenomenon of radiation emission by moving charges,
and that, therefore, we can assume that experimental access
in the near future is entirely realistic.

%
%
\begin{acknowledgments}

The authors acknowledge support by the National Science 
Foundation  (Grant PHY--8555454) 
and by the Missouri Research Board.
The work of E.L.~has been supported by Missouri University of
Science and Technology.

\end{acknowledgments}


\begin{thebibliography}{60}
\expandafter\ifx\csname natexlab\endcsname\relax\def\natexlab#1{#1}\fi
\expandafter\ifx\csname bibnamefont\endcsname\relax
  \def\bibnamefont#1{#1}\fi
\expandafter\ifx\csname bibfnamefont\endcsname\relax
  \def\bibfnamefont#1{#1}\fi
\expandafter\ifx\csname citenamefont\endcsname\relax
  \def\citenamefont#1{#1}\fi
\expandafter\ifx\csname url\endcsname\relax
  \def\url#1{\texttt{#1}}\fi
\expandafter\ifx\csname urlprefix\endcsname\relax\def\urlprefix{URL }\fi
\providecommand{\bibinfo}[2]{#2}
\providecommand{\eprint}[2][]{\url{#2}}

\bibitem[{\citenamefont{Eliezer}(1946)}]{Eliezer1946}
\bibinfo{author}{\bibfnamefont{C.~J.} \bibnamefont{Eliezer}},
  \bibinfo{journal}{Proc. Roy. Soc. (London)} \textbf{\bibinfo{volume}{A187}},
  \bibinfo{pages}{210} (\bibinfo{year}{1946}).

\bibitem[{\citenamefont{Mandl and Skyrme}(1952)}]{MandlSkyrme1952}
\bibinfo{author}{\bibfnamefont{F.}~\bibnamefont{Mandl}} \bibnamefont{and}
  \bibinfo{author}{\bibfnamefont{T.~H.~R.} \bibnamefont{Skyrme}},
  \bibinfo{journal}{Proc. Roy. Soc. London A} \textbf{\bibinfo{volume}{215}},
  \bibinfo{pages}{497} (\bibinfo{year}{1952}).

\bibitem[{\citenamefont{Bell}(2008)}]{Bell2008}
\bibinfo{author}{\bibfnamefont{F.}~\bibnamefont{Bell}},
 \eprint{eprint arXiv:0809.1505v1 [quant-ph]}.

\bibitem[{\citenamefont{Cavanagh}(1952)}]{Cavanagh1952}
\bibinfo{author}{\bibfnamefont{P.~E.} \bibnamefont{Cavanagh}},
  \bibinfo{journal}{Phys. Rev.} \textbf{\bibinfo{volume}{87}},
  \bibinfo{pages}{1131} (\bibinfo{year}{1952}).

\bibitem[{\citenamefont{Theus and Beach}(1957)}]{TheusBeach1957}
\bibinfo{author}{\bibfnamefont{R.~B.} \bibnamefont{Theus}} \bibnamefont{and}
  \bibinfo{author}{\bibfnamefont{L.~A.} \bibnamefont{Beach}},
  \bibinfo{journal}{Phys. Rev.} \textbf{\bibinfo{volume}{106}},
  \bibinfo{pages}{1249} (\bibinfo{year}{1957}).

\bibitem[{\citenamefont{McGie et~al.}(1966)\citenamefont{McGie, Brady, and
  Knox}}]{McGieBradyKnox1966}
\bibinfo{author}{\bibfnamefont{M.~R.} \bibnamefont{McGie}},
  \bibinfo{author}{\bibfnamefont{F.~P.} \bibnamefont{Brady}}, \bibnamefont{and}
  \bibinfo{author}{\bibfnamefont{W.~J.} \bibnamefont{Knox}},
  \bibinfo{journal}{Phys. Rev.} \textbf{\bibinfo{volume}{152}},
  \bibinfo{pages}{1190} (\bibinfo{year}{1966}).

\bibitem[{\citenamefont{Sekhon et~al.}(1988)\citenamefont{Sekhon, Sandhu, and
  Ghumman}}]{SeSaGhu1988}
\bibinfo{author}{\bibfnamefont{G.~S.} \bibnamefont{Sekhon}},
  \bibinfo{author}{\bibfnamefont{B.~S.} \bibnamefont{Sandhu}},
  \bibnamefont{and} \bibinfo{author}{\bibfnamefont{B.~S.}
  \bibnamefont{Ghumman}}, \bibinfo{journal}{Physica C}
  \textbf{\bibinfo{volume}{150}}, \bibinfo{pages}{473} (\bibinfo{year}{1988}).

\bibitem[{\citenamefont{Sandhu et~al.}(1999)\citenamefont{Sandhu, Dewan, Singh,
  and Ghumman}}]{SaDeSiGhu1999}
\bibinfo{author}{\bibfnamefont{B.~S.} \bibnamefont{Sandhu}},
  \bibinfo{author}{\bibfnamefont{R.}~\bibnamefont{Dewan}},
  \bibinfo{author}{\bibfnamefont{B.}~\bibnamefont{Singh}}, \bibnamefont{and}
  \bibinfo{author}{\bibfnamefont{B.~S.} \bibnamefont{Ghumman}},
  \bibinfo{journal}{Phys. Rev. A} \textbf{\bibinfo{volume}{60}},
  \bibinfo{pages}{4600} (\bibinfo{year}{1999}).

\bibitem[{\citenamefont{Saddi et~al.}(2006)\citenamefont{Saddi, Sandhu, and
  Singh}}]{SaSaSi2006}
\bibinfo{author}{\bibfnamefont{M.~B.} \bibnamefont{Saddi}},
  \bibinfo{author}{\bibfnamefont{B.~S.} \bibnamefont{Sandhu}},
  \bibnamefont{and} \bibinfo{author}{\bibfnamefont{B.}~\bibnamefont{Singh}},
  \bibinfo{journal}{Ann. Nucl. Ener.} \textbf{\bibinfo{volume}{33}},
  \bibinfo{pages}{271} (\bibinfo{year}{2006}).

\bibitem[{\citenamefont{Saddi et~al.}(2008)\citenamefont{Saddi, Sing, and
  Sandhu}}]{SaSiSa2008}
\bibinfo{author}{\bibfnamefont{M.~B.} \bibnamefont{Saddi}},
  \bibinfo{author}{\bibfnamefont{B.}~\bibnamefont{Sing}}, \bibnamefont{and}
  \bibinfo{author}{\bibfnamefont{B.~S.} \bibnamefont{Sandhu}},
  \bibinfo{journal}{Nucl. Instrum. Meth. B} \textbf{\bibinfo{volume}{266}},
  \bibinfo{pages}{3309} (\bibinfo{year}{2008}).

\bibitem[{\citenamefont{Melrose}(1972)}]{Mel1972}
\bibinfo{author}{\bibfnamefont{D.~B.} \bibnamefont{Melrose}},
  \bibinfo{journal}{Nuov. Cim.} \textbf{\bibinfo{volume}{7 A}},
  \bibinfo{pages}{669} (\bibinfo{year}{1972}).

\bibitem[{\citenamefont{L\"{o}tstedt et~al.}(2007)\citenamefont{L\"{o}tstedt,
  Jentschura, and Keitel}}]{LoJeKe2007}
\bibinfo{author}{\bibfnamefont{E.}~\bibnamefont{L\"{o}tstedt}},
  \bibinfo{author}{\bibfnamefont{U.~D.} \bibnamefont{Jentschura}},
  \bibnamefont{and} \bibinfo{author}{\bibfnamefont{C.~H.}
  \bibnamefont{Keitel}}, \bibinfo{journal}{Phys. Rev. Lett.}
  \textbf{\bibinfo{volume}{98}}, \bibinfo{pages}{043002}
  (\bibinfo{year}{2007}).

\bibitem[{\citenamefont{Reiss and Eberly}(1966)}]{ReEb1966}
\bibinfo{author}{\bibfnamefont{H.~R.} \bibnamefont{Reiss}} \bibnamefont{and}
  \bibinfo{author}{\bibfnamefont{J.~H.} \bibnamefont{Eberly}},
  \bibinfo{journal}{Phys. Rev.} \textbf{\bibinfo{volume}{151}},
  \bibinfo{pages}{1058} (\bibinfo{year}{1966}).

\bibitem[{\citenamefont{Brodin et~al.}(2008)\citenamefont{Brodin, Marklund,
  Bingham, Collier, and Evans}}]{BroMarBinColEva2008}
\bibinfo{author}{\bibfnamefont{G.}~\bibnamefont{Brodin}},
  \bibinfo{author}{\bibfnamefont{M.}~\bibnamefont{Marklund}},
  \bibinfo{author}{\bibfnamefont{R.}~\bibnamefont{Bingham}},
  \bibinfo{author}{\bibfnamefont{J.}~\bibnamefont{Collier}}, \bibnamefont{and}
  \bibinfo{author}{\bibfnamefont{R.~G.} \bibnamefont{Evans}},
  \bibinfo{journal}{Class. Quantum Grav.} \textbf{\bibinfo{volume}{25}},
  \bibinfo{pages}{145005} (\bibinfo{year}{2008}).

\bibitem[{\citenamefont{Thirolf et~al.}(2009)\citenamefont{Thirolf, Habs,
  Henig, Jung, Kiefer, Lang, Schreiber, Maia, Schaller, Sch??tzhold
  et~al.}}]{Thirolfetal2009}
\bibinfo{author}{\bibfnamefont{P.~G.} \bibnamefont{Thirolf}},
  \bibinfo{author}{\bibfnamefont{D.}~\bibnamefont{Habs}},
  \bibinfo{author}{\bibfnamefont{A.}~\bibnamefont{Henig}},
  \bibinfo{author}{\bibfnamefont{D.}~\bibnamefont{Jung}},
  \bibinfo{author}{\bibfnamefont{D.}~\bibnamefont{Kiefer}},
  \bibinfo{author}{\bibfnamefont{C.}~\bibnamefont{Lang}},
  \bibinfo{author}{\bibfnamefont{J.}~\bibnamefont{Schreiber}},
  \bibinfo{author}{\bibfnamefont{C.}~\bibnamefont{Maia}},
  \bibinfo{author}{\bibfnamefont{G.}~\bibnamefont{Schaller}},
  \bibinfo{author}{\bibfnamefont{R.}~\bibnamefont{Sch\"{u}tzhold}},
  \bibnamefont{et~al.}, \bibinfo{journal}{Eur. Phys. J. D} \textbf{\bibinfo{volume}{55}},
  \bibinfo{pages}{379}
  (\bibinfo{year}{2009}).

\bibitem[{\citenamefont{Sch\"{u}tzhold
  et~al.}(2006)\citenamefont{Sch\"{u}tzhold, Schaller, and Habs}}]{ScScHa2006}
\bibinfo{author}{\bibfnamefont{R.}~\bibnamefont{Sch\"{u}tzhold}},
  \bibinfo{author}{\bibfnamefont{G.}~\bibnamefont{Schaller}}, \bibnamefont{and}
  \bibinfo{author}{\bibfnamefont{D.}~\bibnamefont{Habs}},
  \bibinfo{journal}{Phys. Rev. Lett.} \textbf{\bibinfo{volume}{97}},
  \bibinfo{eid}{121302} (\bibinfo{year}{2006}).

\bibitem[{\citenamefont{Sch\"{u}tzhold
  et~al.}(2008)\citenamefont{Sch\"{u}tzhold, Schaller, and Habs}}]{ScScHa2008}
\bibinfo{author}{\bibfnamefont{R.}~\bibnamefont{Sch\"{u}tzhold}},
  \bibinfo{author}{\bibfnamefont{G.}~\bibnamefont{Schaller}}, \bibnamefont{and}
  \bibinfo{author}{\bibfnamefont{D.}~\bibnamefont{Habs}},
  \bibinfo{journal}{Phys. Rev. Lett.} \textbf{\bibinfo{volume}{100}},
  \bibinfo{eid}{091301} (\bibinfo{year}{2008}).

\bibitem[{\citenamefont{Sch\"{u}tzhold and Maia}(2009)}]{ScMai2009}
\bibinfo{author}{\bibfnamefont{R.}~\bibnamefont{Sch\"{u}tzhold}}
  \bibnamefont{and} \bibinfo{author}{\bibfnamefont{C.}~\bibnamefont{Maia}},
  \bibinfo{journal}{Eur. Phys. J. D}  \textbf{\bibinfo{volume}{55}},
  \bibinfo{pages}{375}
  (\bibinfo{year}{2009}).

\bibitem[{\citenamefont{Wootters}(1998)}]{Wootters1998}
\bibinfo{author}{\bibfnamefont{W.~K.} \bibnamefont{Wootters}},
  \bibinfo{journal}{Phys. Rev. Lett.} \textbf{\bibinfo{volume}{80}},
  \bibinfo{pages}{2245} (\bibinfo{year}{1998}).

\bibitem[{\citenamefont{Cirone}(2005)}]{Cirone2005269}
\bibinfo{author}{\bibfnamefont{M.~A.} \bibnamefont{Cirone}},
  \bibinfo{journal}{Phys. Lett. A} \textbf{\bibinfo{volume}{339}},
  \bibinfo{pages}{269 } (\bibinfo{year}{2005}).

\bibitem[{\citenamefont{Baier et~al.}(1981)\citenamefont{Baier, Fadin, Khoze,
  and Kuraev}}]{Baier1981293}
\bibinfo{author}{\bibfnamefont{V.~N.} \bibnamefont{Baier}},
  \bibinfo{author}{\bibfnamefont{V.~S.} \bibnamefont{Fadin}},
  \bibinfo{author}{\bibfnamefont{V.~A.} \bibnamefont{Khoze}}, \bibnamefont{and}
  \bibinfo{author}{\bibfnamefont{E.~A.} \bibnamefont{Kuraev}},
  \bibinfo{journal}{Phys. Rep.} \textbf{\bibinfo{volume}{78}},
  \bibinfo{pages}{293 } (\bibinfo{year}{1981}).

\bibitem[{\citenamefont{Altman and Quarles}(1985)}]{AltQuar1985}
\bibinfo{author}{\bibfnamefont{J.~C.} \bibnamefont{Altman}} \bibnamefont{and}
  \bibinfo{author}{\bibfnamefont{C.~A.} \bibnamefont{Quarles}},
  \bibinfo{journal}{Phys. Rev. A} \textbf{\bibinfo{volume}{31}},
  \bibinfo{pages}{R2744} (\bibinfo{year}{1985}).

\bibitem[{\citenamefont{V\'eniard et~al.}(1987)\citenamefont{V\'eniard,
  Gavrila, and Maquet}}]{VenGavMaq1987}
\bibinfo{author}{\bibfnamefont{V.}~\bibnamefont{V\'eniard}},
  \bibinfo{author}{\bibfnamefont{M.}~\bibnamefont{Gavrila}}, \bibnamefont{and}
  \bibinfo{author}{\bibfnamefont{A.}~\bibnamefont{Maquet}},
  \bibinfo{journal}{Phys. Rev. A} \textbf{\bibinfo{volume}{35}},
  \bibinfo{pages}{R448} (\bibinfo{year}{1987}).

\bibitem[{\citenamefont{Hippler}(1991)}]{Hippler1991}
\bibinfo{author}{\bibfnamefont{R.}~\bibnamefont{Hippler}},
  \bibinfo{journal}{Phys. Rev. Lett.} \textbf{\bibinfo{volume}{66}},
  \bibinfo{pages}{2197} (\bibinfo{year}{1991}).

\bibitem[{\citenamefont{Kahler et~al.}(1992)\citenamefont{Kahler, Liu, and
  Quarles}}]{KahLiuQuar1992}
\bibinfo{author}{\bibfnamefont{D.~L.} \bibnamefont{Kahler}},
  \bibinfo{author}{\bibfnamefont{J.}~\bibnamefont{Liu}}, \bibnamefont{and}
  \bibinfo{author}{\bibfnamefont{C.~A.} \bibnamefont{Quarles}},
  \bibinfo{journal}{Phys. Rev. Lett.} \textbf{\bibinfo{volume}{68}},
  \bibinfo{pages}{1690} (\bibinfo{year}{1992}).

\bibitem[{\citenamefont{Korol}(1997)}]{Korol1997}
\bibinfo{author}{\bibfnamefont{A.~V.} \bibnamefont{Korol}},
  \bibinfo{journal}{J. Phys. B} \textbf{\bibinfo{volume}{30}},
  \bibinfo{pages}{413} (\bibinfo{year}{1997}).

\bibitem[{\citenamefont{Dondera and Florescu}(1998)}]{DonFlo1998}
\bibinfo{author}{\bibfnamefont{M.}~\bibnamefont{Dondera}} \bibnamefont{and}
  \bibinfo{author}{\bibfnamefont{V.}~\bibnamefont{Florescu}},
  \bibinfo{journal}{Phys. Rev. A} \textbf{\bibinfo{volume}{58}},
  \bibinfo{pages}{2016} (\bibinfo{year}{1998}).

\bibitem[{\citenamefont{Krylovetski{\u \i}
  et~al.}(2002)\citenamefont{Krylovetski{\u \i}, Manakov, Marmo, and
  Starace}}]{KryMaMaSta2002}
\bibinfo{author}{\bibfnamefont{A.~A.} \bibnamefont{Krylovetski{\u \i}}},
  \bibinfo{author}{\bibfnamefont{N.~L.} \bibnamefont{Manakov}},
  \bibinfo{author}{\bibfnamefont{S.~I.} \bibnamefont{Marmo}}, \bibnamefont{and}
  \bibinfo{author}{\bibfnamefont{A.~F.} \bibnamefont{Starace}},
  \bibinfo{journal}{Zh. \'{E}ksp. Teor. Fiz.} \textbf{\bibinfo{volume}{122}},
  \bibinfo{pages}{1168} (\bibinfo{year}{2002}), \bibinfo{note}{[JETP
  \textbf{95}, 1006 (2002)]}.

\bibitem[{\citenamefont{Korol and Solovjev}(2006)}]{Korol2006}
\bibinfo{author}{\bibfnamefont{A.~V.} \bibnamefont{Korol}} \bibnamefont{and}
  \bibinfo{author}{\bibfnamefont{I.~A.} \bibnamefont{Solovjev}},
  \bibinfo{journal}{Rad. Phys. Chem.} \textbf{\bibinfo{volume}{75}},
  \bibinfo{pages}{1346} (\bibinfo{year}{2006}).

\bibitem[{\citenamefont{Zhukovski\u\i{} and Nikitina}(1973)}]{ZhuNi1973}
\bibinfo{author}{\bibfnamefont{V.~C.} \bibnamefont{Zhukovski\u\i{}}}
  \bibnamefont{and} \bibinfo{author}{\bibfnamefont{N.~S.}
  \bibnamefont{Nikitina}}, \bibinfo{journal}{Zh. \'{E}ksp. Teor. Fiz.}
  \textbf{\bibinfo{volume}{64}}, \bibinfo{pages}{1169} (\bibinfo{year}{1973}),
  \bibinfo{note}{[Sov. Phys.-JETP \textbf{37}, 595 (1973)]}.

\bibitem[{\citenamefont{Sokolov et~al.}(1976)\citenamefont{Sokolov,
  Voloshchenko, Zhukovskii, and Pavlenko}}]{SoVoletal1976}
\bibinfo{author}{\bibfnamefont{A.~A.} \bibnamefont{Sokolov}},
  \bibinfo{author}{\bibfnamefont{A.~M.} \bibnamefont{Voloshchenko}},
  \bibinfo{author}{\bibfnamefont{V.~C.} \bibnamefont{Zhukovskii}},
  \bibnamefont{and} \bibinfo{author}{\bibfnamefont{{\relax Yu}.~G.}
  \bibnamefont{Pavlenko}}, \bibinfo{journal}{Izv. Vyssh. Uchebn. Zaved., Fiz.}
  \textbf{\bibinfo{volume}{19}}, \bibinfo{pages}{46} (\bibinfo{year}{1976}),
  \bibinfo{note}{[Rus. Phys. J. \textbf{19}, 1139 (1976)]}.

\bibitem[{\citenamefont{Fomin and Kholodov}(2003)}]{FoKho2003}
\bibinfo{author}{\bibfnamefont{P.~I.} \bibnamefont{Fomin}} \bibnamefont{and}
  \bibinfo{author}{\bibfnamefont{R.~I.} \bibnamefont{Kholodov}},
  \bibinfo{journal}{Zh. \'{E}ksp. Teor. Fiz.} \textbf{\bibinfo{volume}{123}},
  \bibinfo{pages}{356} (\bibinfo{year}{2003}), \bibinfo{note}{[JETP
  \textbf{96}, 315 (2003)]}.

\bibitem[{\citenamefont{Morozov and Ritus}(1975)}]{MoRi1974}
\bibinfo{author}{\bibfnamefont{D.~A.} \bibnamefont{Morozov}} \bibnamefont{and}
  \bibinfo{author}{\bibfnamefont{V.~I.} \bibnamefont{Ritus}},
  \bibinfo{journal}{Nucl. Phys. B} \textbf{\bibinfo{volume}{86}},
  \bibinfo{pages}{309} (\bibinfo{year}{1975}).

\bibitem[{\citenamefont{L\"{o}tstedt and
  Jentschura}(2009{\natexlab{a}})}]{LoJe2009_2}
\bibinfo{author}{\bibfnamefont{E.}~\bibnamefont{L\"{o}tstedt}}
  \bibnamefont{and} \bibinfo{author}{\bibfnamefont{U.~D.}
  \bibnamefont{Jentschura}}, 
\bibinfo{journal}{Phys. Rev. Lett.} 
\textbf{\bibinfo{volume}{103}},
 \bibinfo{pages}{110404} 
(\bibinfo{year}{2009}).

\bibitem[{\citenamefont{Ritus}(1979)}]{Ri1985}
\bibinfo{author}{\bibfnamefont{V.~I.} \bibnamefont{Ritus}},
  \bibinfo{journal}{Trud. Ord. Len. Fiz. Inst. im. P. N. Leb. Akad. Nauk SSSR}
  \textbf{\bibinfo{volume}{111}}, \bibinfo{pages}{5} (\bibinfo{year}{1979}),
  \bibinfo{note}{[J. Rus. Laser Res. \textbf{6}, 497 (1985)]}.

\bibitem[{\citenamefont{Reiss}(1962)}]{Re1962}
\bibinfo{author}{\bibfnamefont{H.~R.} \bibnamefont{Reiss}},
  \bibinfo{journal}{J. Math. Phys.} \textbf{\bibinfo{volume}{3}},
  \bibinfo{pages}{59} (\bibinfo{year}{1962}).

\bibitem[{\citenamefont{Korsch et~al.}(2006)\citenamefont{Korsch, Klumpp, and
  Witthaut}}]{KoKlWi2006}
\bibinfo{author}{\bibfnamefont{H.~J.} \bibnamefont{Korsch}},
  \bibinfo{author}{\bibfnamefont{A.}~\bibnamefont{Klumpp}}, \bibnamefont{and}
  \bibinfo{author}{\bibfnamefont{D.}~\bibnamefont{Witthaut}},
  \bibinfo{journal}{J. Phys. A} \textbf{\bibinfo{volume}{39}},
  \bibinfo{pages}{14947} (\bibinfo{year}{2006}).

\bibitem[{\citenamefont{Roshchupkin}(1985)}]{Ro1985}
\bibinfo{author}{\bibfnamefont{S.~P.} \bibnamefont{Roshchupkin}},
  \bibinfo{journal}{Yad. Fiz.} \textbf{\bibinfo{volume}{41}},
  \bibinfo{pages}{1244} (\bibinfo{year}{1985}), \bibinfo{note}{[Sov. J. Nucl.
  Phys. \textbf{41}, 796 (1985)]}.

\bibitem[{\citenamefont{Nikishov and Ritus}(1964)}]{NiRi1964}
\bibinfo{author}{\bibfnamefont{A.~I.} \bibnamefont{Nikishov}} \bibnamefont{and}
  \bibinfo{author}{\bibfnamefont{V.~I.} \bibnamefont{Ritus}},
  \bibinfo{journal}{Zh. \'{E}ksp. Teor. Fiz.} \textbf{\bibinfo{volume}{46}},
  \bibinfo{pages}{776} (\bibinfo{year}{1964}), \bibinfo{note}{[Sov. Phys. JETP
  \textbf{19}, 529 (1964)]}.

\bibitem[{\citenamefont{Harvey et~al.}(2009)\citenamefont{Harvey, Heinzl, and
  Ilderton}}]{HarHeinIld2009}
\bibinfo{author}{\bibfnamefont{C.}~\bibnamefont{Harvey}},
  \bibinfo{author}{\bibfnamefont{T.}~\bibnamefont{Heinzl}}, \bibnamefont{and}
  \bibinfo{author}{\bibfnamefont{A.}~\bibnamefont{Ilderton}},
  \bibinfo{journal}{Phys. Rev. A} \textbf{\bibinfo{volume}{79}},
  \bibinfo{eid}{063407} (\bibinfo{year}{2009}).

\bibitem[{\citenamefont{Panek et~al.}(2002)\citenamefont{Panek, Kami{\'n}ski,
  and Ehlotzky}}]{PaKaEh2002}
\bibinfo{author}{\bibfnamefont{P.}~\bibnamefont{Panek}},
  \bibinfo{author}{\bibfnamefont{J.~Z.} \bibnamefont{Kami{\'n}ski}},
  \bibnamefont{and} \bibinfo{author}{\bibfnamefont{F.}~\bibnamefont{Ehlotzky}},
  \bibinfo{journal}{Phys. Rev. A} \textbf{\bibinfo{volume}{65}},
  \bibinfo{pages}{022712} (\bibinfo{year}{2002}).

\bibitem[{\citenamefont{Bamber et~al.}(1999)\citenamefont{Bamber, Boege,
  Koffas, Kotseroglou, Melissinos, Meyerhofer, Reis, Ragg, Bula, McDonald
  et~al.}}]{BaBoetal2004}
\bibinfo{author}{\bibfnamefont{C.}~\bibnamefont{Bamber}},
  \bibinfo{author}{\bibfnamefont{S.~J.} \bibnamefont{Boege}},
  \bibinfo{author}{\bibfnamefont{T.}~\bibnamefont{Koffas}},
  \bibinfo{author}{\bibfnamefont{T.}~\bibnamefont{Kotseroglou}},
  \bibinfo{author}{\bibfnamefont{A.~C.} \bibnamefont{Melissinos}},
  \bibinfo{author}{\bibfnamefont{D.~D.} \bibnamefont{Meyerhofer}},
  \bibinfo{author}{\bibfnamefont{D.~A.} \bibnamefont{Reis}},
  \bibinfo{author}{\bibfnamefont{W.}~\bibnamefont{Ragg}},
  \bibinfo{author}{\bibfnamefont{C.}~\bibnamefont{Bula}},
  \bibinfo{author}{\bibfnamefont{K.~T.} \bibnamefont{McDonald}},
  \bibnamefont{et~al.}, \bibinfo{journal}{Phys. Rev. D}
  \textbf{\bibinfo{volume}{60}}, \bibinfo{pages}{092004}
  (\bibinfo{year}{1999}).

\bibitem[{\citenamefont{L\"{o}tstedt and
  Jentschura}(2009{\natexlab{b}})}]{LoJe2009}
\bibinfo{author}{\bibfnamefont{E.}~\bibnamefont{L\"{o}tstedt}}
  \bibnamefont{and} \bibinfo{author}{\bibfnamefont{U.~D.}
  \bibnamefont{Jentschura}}, \bibinfo{journal}{Phys. Rev. E}
  \textbf{\bibinfo{volume}{79}}, \bibinfo{pages}{026707}
  (\bibinfo{year}{2009}{\natexlab{b}}).

\bibitem[{\citenamefont{Becker and Mitter}(1976)}]{BeMi1976}
\bibinfo{author}{\bibfnamefont{W.}~\bibnamefont{Becker}} \bibnamefont{and}
  \bibinfo{author}{\bibfnamefont{H.}~\bibnamefont{Mitter}},
  \bibinfo{journal}{J. Phys. A} \textbf{\bibinfo{volume}{9}},
  \bibinfo{pages}{2171} (\bibinfo{year}{1976}).

\bibitem[{\citenamefont{Schnez et~al.}(2007)\citenamefont{Schnez, L\"{o}tstedt,
  Jentschura, and Keitel}}]{SchLoJeKe2007}
\bibinfo{author}{\bibfnamefont{S.}~\bibnamefont{Schnez}},
  \bibinfo{author}{\bibfnamefont{E.}~\bibnamefont{L\"{o}tstedt}},
  \bibinfo{author}{\bibfnamefont{U.~D.} \bibnamefont{Jentschura}},
  \bibnamefont{and} \bibinfo{author}{\bibfnamefont{C.~H.}
  \bibnamefont{Keitel}}, \bibinfo{journal}{Phys. Rev. A}
  \textbf{\bibinfo{volume}{75}}, \bibinfo{eid}{053412} (\bibinfo{year}{2007}).

\bibitem[{\citenamefont{Jentschura and
  Surzhykov}(2008)}]{jentschurasurzhykov2008}
\bibinfo{author}{\bibfnamefont{U.~D.} \bibnamefont{Jentschura}}
  \bibnamefont{and}
  \bibinfo{author}{\bibfnamefont{A.}~\bibnamefont{Surzhykov}},
  \bibinfo{journal}{Phys. Rev. A} \textbf{\bibinfo{volume}{77}},
  \bibinfo{eid}{042507} (\bibinfo{year}{2008}).

\bibitem[{\citenamefont{Jentschura}(2007)}]{UDJ2007}
\bibinfo{author}{\bibfnamefont{U.~D.} \bibnamefont{Jentschura}},
  \bibinfo{journal}{J. Phys. A} \textbf{\bibinfo{volume}{40}},
  \bibinfo{pages}{F223} (\bibinfo{year}{2007}).

\bibitem[{\citenamefont{Jentschura}(2008)}]{UDJ2008}
\bibinfo{author}{\bibfnamefont{U.~D.} \bibnamefont{Jentschura}},
  \bibinfo{journal}{J. Phys. A} \textbf{\bibinfo{volume}{41}},
  \bibinfo{pages}{155307} (\bibinfo{year}{2008}).

\bibitem[{\citenamefont{Jentschura}(2009)}]{UDJ2009}
\bibinfo{author}{\bibfnamefont{U.~D.} \bibnamefont{Jentschura}},
  \bibinfo{journal}{Phys. Rev. A} \textbf{\bibinfo{volume}{79}},
  \bibinfo{eid}{022510} (\bibinfo{year}{2009}).

\bibitem[{\citenamefont{Amaro et~al.}(2009)\citenamefont{Amaro, Santos,
  Parente, Surzhykov, and Indelicato}}]{amaroetal2009}
\bibinfo{author}{\bibfnamefont{P.}~\bibnamefont{Amaro}},
  \bibinfo{author}{\bibfnamefont{J.~P.} \bibnamefont{Santos}},
  \bibinfo{author}{\bibfnamefont{F.}~\bibnamefont{Parente}},
  \bibinfo{author}{\bibfnamefont{A.}~\bibnamefont{Surzhykov}},
  \bibnamefont{and}
  \bibinfo{author}{\bibfnamefont{P.}~\bibnamefont{Indelicato}},
  \bibinfo{journal}{Phys. Rev. A} \textbf{\bibinfo{volume}{79}},
  \bibinfo{eid}{062504} (\bibinfo{year}{2009}).

\bibitem[{\citenamefont{Jauch and Rohrlich}(1976)}]{JauchRohr1976}
\bibinfo{author}{\bibfnamefont{J.~M.} \bibnamefont{Jauch}} \bibnamefont{and}
  \bibinfo{author}{\bibfnamefont{F.}~\bibnamefont{Rohrlich}},
  \emph{\bibinfo{title}{The theory of photons and electrons}}
  (\bibinfo{publisher}{Addison-Wesley publishing company},
  \bibinfo{address}{Reading, Massachusetts}, \bibinfo{year}{1976}),
  \bibinfo{edition}{2nd} ed.

\bibitem[{\citenamefont{Ram and Wang}(1971)}]{RamWang1971}
\bibinfo{author}{\bibfnamefont{M.}~\bibnamefont{Ram}} \bibnamefont{and}
  \bibinfo{author}{\bibfnamefont{P.~Y.} \bibnamefont{Wang}},
  \bibinfo{journal}{Phys. Rev. Lett.} \textbf{\bibinfo{volume}{26}},
  \bibinfo{pages}{476} (\bibinfo{year}{1971}); \bibinfo{note}{erratum: Phys.
  Rev. Lett. {\bf 26}, 1210 (1971)}.

\bibitem[{\citenamefont{Gould}(1984)}]{Gould1984}
\bibinfo{author}{\bibfnamefont{R.~J.} \bibnamefont{Gould}},
  \bibinfo{journal}{Astrophysical J.} \textbf{\bibinfo{volume}{285}},
  \bibinfo{pages}{275} (\bibinfo{year}{1984}).

\bibitem[{\citenamefont{Radha and Thunga}(1961)}]{RaThu1961}
\bibinfo{author}{\bibfnamefont{T.~K.} \bibnamefont{Radha}} \bibnamefont{and}
  \bibinfo{author}{\bibfnamefont{R.}~\bibnamefont{Thunga}},
  \bibinfo{journal}{Z. Phys.} \textbf{\bibinfo{volume}{161}},
  \bibinfo{pages}{20} (\bibinfo{year}{1961}).

\bibitem[{\citenamefont{Carrassi and Passatore}(1960)}]{CarPas1960}
\bibinfo{author}{\bibfnamefont{M.}~\bibnamefont{Carrassi}} \bibnamefont{and}
  \bibinfo{author}{\bibfnamefont{G.}~\bibnamefont{Passatore}},
  \bibinfo{journal}{Nuov. Cim.} \textbf{\bibinfo{volume}{16}},
  \bibinfo{pages}{1835} (\bibinfo{year}{1960}).

\bibitem[{\citenamefont{Radtke et~al.}(2008{\natexlab{a}})\citenamefont{Radtke,
  Surzhykov, and Fritzsche}}]{radtke2008}
\bibinfo{author}{\bibfnamefont{T.}~\bibnamefont{Radtke}},
  \bibinfo{author}{\bibfnamefont{A.}~\bibnamefont{Surzhykov}},
  \bibnamefont{and}
  \bibinfo{author}{\bibfnamefont{S.}~\bibnamefont{Fritzsche}},
  \bibinfo{journal}{Phys. Rev. A} \textbf{\bibinfo{volume}{77}},
  \bibinfo{eid}{022507}
  (\bibinfo{year}{2008}{\natexlab{a}}).

\bibitem[{\citenamefont{Radtke et~al.}(2008{\natexlab{b}})\citenamefont{Radtke,
  Surzhykov, and Fritzsche}}]{Radke2008_2}
\bibinfo{author}{\bibfnamefont{T.}~\bibnamefont{Radtke}},
  \bibinfo{author}{\bibfnamefont{A.}~\bibnamefont{Surzhykov}},
  \bibnamefont{and}
  \bibinfo{author}{\bibfnamefont{S.}~\bibnamefont{Fritzsche}},
  \bibinfo{journal}{Eur. Phys. J. D} \textbf{\bibinfo{volume}{49}},
  \bibinfo{pages}{7} (\bibinfo{year}{2008}{\natexlab{b}}).

\bibitem[{\citenamefont{Maiorova et~al.}(2009)\citenamefont{Maiorova,
  Surzhykov, Tashenov, Shabaev, Fritzsche, Plunien, and
  St{\"o}hlker}}]{MaiSuretal2009}
\bibinfo{author}{\bibfnamefont{A.~V.} \bibnamefont{Maiorova}},
  \bibinfo{author}{\bibfnamefont{A.}~\bibnamefont{Surzhykov}},
  \bibinfo{author}{\bibfnamefont{S.}~\bibnamefont{Tashenov}},
  \bibinfo{author}{\bibfnamefont{V.~M.} \bibnamefont{Shabaev}},
  \bibinfo{author}{\bibfnamefont{S.}~\bibnamefont{Fritzsche}},
  \bibinfo{author}{\bibfnamefont{G.}~\bibnamefont{Plunien}}, \bibnamefont{and}
  \bibinfo{author}{\bibfnamefont{T.}~\bibnamefont{St{\"o}hlker}},
  \bibinfo{journal}{J. Phys. B} \textbf{\bibinfo{volume}{42}},
  \bibinfo{pages}{125003} (\bibinfo{year}{2009}).

\bibitem[{\citenamefont{Radtke et~al.}(2006)\citenamefont{Radtke, Fritzsche,
  and Surzhykov}}]{radtkeetal2006}
\bibinfo{author}{\bibfnamefont{T.}~\bibnamefont{Radtke}},
  \bibinfo{author}{\bibfnamefont{S.}~\bibnamefont{Fritzsche}},
  \bibnamefont{and}
  \bibinfo{author}{\bibfnamefont{A.}~\bibnamefont{Surzhykov}},
  \bibinfo{journal}{Phys. Rev. A} \textbf{\bibinfo{volume}{74}},
  \bibinfo{eid}{032709} (\bibinfo{year}{2006}).

\bibitem[{\citenamefont{Blum}(1996)}]{Blum1996}
\bibinfo{author}{\bibfnamefont{K.}~\bibnamefont{Blum}},
  \emph{\bibinfo{title}{Density matrix theory and applications}}
  (\bibinfo{publisher}{Plenum Press}, \bibinfo{address}{New York},
  \bibinfo{year}{1996}), \bibinfo{edition}{2nd} ed.

\end{thebibliography}

\end{document}